\documentclass[10pt,journal]{IEEEtran}\IEEEoverridecommandlockouts
\usepackage{amssymb}
\usepackage{amsmath}
\usepackage{amsfonts}
\usepackage{graphicx}
\usepackage{subfigure}
\usepackage{algorithm}
\usepackage{algorithmic}
\usepackage{tabularx}
\usepackage{cite}
\usepackage{stfloats}
\usepackage{color}
\definecolor{orange}{RGB}{255,127,0}
\begin{document}
%Reducing the Computational Complexity of Multiuser Hybrid Analog/Digital Beamforming in Large-Scale Antenna Systems
\title{QoE-Oriented Resource Allocation for 360-degree Video Transmission over Heterogeneous Networks}

\author{\IEEEauthorblockN{Wei Huang\IEEEauthorrefmark{1}\thanks{This paper has been submitted to Digital Signal Processing.}, Lianghui Ding\IEEEauthorrefmark{1}, Hung-Yu Wei\IEEEauthorrefmark{3}~\IEEEmembership{Member,~IEEE},\\Jenq-Neng Hwang\IEEEauthorrefmark{2}~\IEEEmembership{Fellow,~IEEE}, Yiling Xu\IEEEauthorrefmark{1}, and Wenjun Zhang\IEEEauthorrefmark{1}}~\IEEEmembership{Fellow,~IEEE},

\IEEEauthorblockA{\IEEEauthorrefmark{1}Cooperative Media Network Innovation Center, Shanghai Jiao Tong University, Shanghai, China}

\IEEEauthorblockA{\IEEEauthorrefmark{2}Department of Electrical Engineering, University of Washington, Seattle WA, USA}

\IEEEauthorblockA{\IEEEauthorrefmark{3}Department of Electrical Engineering,  National Taiwan University, Taipei, Taiwan}

}

\maketitle

\begin{abstract}

Immersive media streaming, especially virtual reality (VR)/360-degree video streaming which is very bandwidth demanding, has become more and more popular due to the rapid growth of the multimedia and networking deployments. To better explore the usage of resource and achieve better quality of experience (QoE) perceived by users, this paper develops an application-layer scheme to jointly exploit the available bandwidth from the LTE and Wi-Fi networks in 360-degree video streaming. This newly proposed scheme and the corresponding solution algorithms utilize the saliency of video, prediction of users' view and the status information of users to obtain an optimal association of the users with different Wi-Fi access points (APs) for maximizing the system's utility. Besides, a novel buffer strategy is proposed to mitigate the influence of short-time prediction problem for transmitting 360-degree videos in time-varying networks. The promising performance and low complexity of the proposed scheme and algorithms are validated in simulations with various 360-degree videos.

\end{abstract}

\begin{IEEEkeywords}
Virtual reality (VR)/360-degree video; quality of experience (QoE); field of view (FoV); saliency; resource allocation; multiple access technology; buffer management.
\end{IEEEkeywords}

\section{Introduction}
\IEEEPARstart{R}{ecently}, immersive media has gained increasing popularity, including 360-degree/VR videos and argument reality/holograms, since they can provide people personalized  and immersive experience. Especially 360-degree/VR videos can now  be easily perceived by users through head-mounted displays (HMDs). Most 360-degree/VR videos have a resolution higher than 4K to provide a real immersive feel, therefore the desire for better immersion and presence has placed new demands on the network in terms of its quality and performance, especially the quality of experience (QoE) perceived by users. However, bandwidth requirements will become increasingly imperative correspondingly for a high-quality virtual reality experience. According to the data from \cite{Cisco1}, VR traffic is poised to grow 20-fold by 2021. Providers need to take a note of the new demands and enhance the experience of users with limited bandwidth.

Tile-based videos have been widely used in immersive media to enable the adaptive video transmission  based on user's region of interests (ROIs). Typically, in 360-degree/VR videos, users can only see parts of the video at a certain time. If the server transmits the whole video, most bandwidth will be wasted to transmit the video not visible to users. A tile-based method spatially partitions  a 360-degree/VR video into multiple segments, which are called tiles. Some papers have proposed methods to transmit the tiles according to the user's field of view (FoV) \cite{Bell,prediction,arxiv,QEC,VR1,FoV,VR2}. By predicting the users' behavior and getting the FoV of users, only tiles in the FoV are transmitted with high quality while minimizing the quality of the rest of the video to save bandwidth. However, the existing algorithms only consider a single user in transmission and cannot be easily extended to wireless VR transmission with multiple users, since the server needs to consider the rate allocation on users and on tiles simultaneously. Besides, they use a constant number of tiles in the FoV, ignoring the fact  that the number of tiles corresponding to the FoV will change depending on the viewpoint \cite{arxiv}. Moreover, users will pay more attention on parts of the 360-degree videos corresponding to their FoV. These features in 360-degree video will drastically change the transmission strategies.

Researchers have investigated the saliency in 360-degree videos~\cite{saliency1,saliency2,saliency3}, which is used to denote the most probable areas in a video an  average person will look at. Furthermore, it is pointed out that average motion in  a  360-degree video is less than that for a regular video\cite{motion}. These characteristics  can be highly beneficial for reducing the bandwidth consumption in 360-degree video transmission, nonetheless, few papers investigate how to utilize these characteristics collectively with resource allocation in wireless VR transmission.

The evolution of 5G technology has contributed to providing massive improvements for 360-degree videos in terms of bandwidth and reliability. A lot of researchers have tried to better exploit the bandwidth by utilizing the heterogeneous LTE and WLAN multi-radio networks. They allow users access to such heterogeneous bandwidth by operating the multi-radio interfaces simultaneously. Typical solutions include LTE-WLAN aggregation technology \cite{LWA1,LWA2,LWA3} and software-defined networking (SDN) used to help further exploit the heterogeneous resource flexibly \cite{SDN1,SDN2}. However, applying these technologies to 360-degree videos requires solving a series of problems, such as jointly optimizing the spatial and temporal domains, controlling network association with multiple users, combining the saliency and users' field of view, and achieving low complexity. Some researchers have investigated the 360-degree video streaming with these new technologies \cite{SDNVR1,SDNVR2,SDNVR3,SDNVR4}. Nonetheless, their works are based on single user and detailed resource allocation schemes are missing.

%Software-defined networking (SDN), which has gained a great popularity recently, achieves high efficiency through decoupling the control plane from the data plane \cite{SDN1,SDN2}. Consequently, it can utilize the heterogeneous networks to flexibly and dynamically optimize network flows based on global network information, such as technologies allow user access to the LTE and Wi-Fi networks simultaneously \cite{LWA1,LWA2,LWA3}.  SDN-based systems for 360-degree video streaming have been proposed \cite{SDNVR1,SDNVR2,SDNVR3,SDNVR4}, e.g., one investigates the performance of 360-degree videos transmission with SDN-based DASH system \cite{SDNVR3}.

%Nonetheless, detailed resource allocation schemes are missing in these works.

The aforementioned technologies and methods rely on the prior information of channel states. If the feedback fails to estimate the fluctuation in time-varying networks, users may see a frozen/blank screen and wait for the arrival of the next frame, resulting in a big drop of QoE. Playback buffer can tackle this problem by storing videos in advance \cite{buffer1,buffer2,buffer3}. However, the scheme is not suitable for FoV-driven 360-degree videos since it is difficult to predict the FoV for a long time in the buffer, the server needs to transmit the whole video for those poorly predicted frames. Nonetheless, a short buffer can cause frequent re-buffering events and result in the video pause and low QoE when the client's playback buffer goes empty~\cite{re-buffering}. On the other hand, a long buffer can incur poor prediction results and also result in low QoE or wasted bandwidth.

In this paper, we propose an application-layer (APP-layer) resource allocation scheme for 360-degree/VR video transmissions over multi-RAT systems with multiple users. We jointly consider how saliency and FoV positions influence the tile-based 360-degree videos in streaming. More specifically, we propose algorithms to decide which user should be connected to which Wi-Fi AP, and choose appropriate transmission rates for each tile of each video, so that the overall system QoE (utility) can be maximized. Additionally, to address the buffer problem of FoV-driven 360-degree video networking schemes, we propose a novel buffer strategy to achieve a good tradeoff between the video quality and the buffer length. The strategy can be combined with our proposed resource allocation algorithms and achieve better QoE for time-varying networks. In summary, our work makes contributions as follows:
\begin{enumerate}
\item We propose a new 360-degree video transmission scheme, which combines saliency and FoV in wireless multi-RAT networks, to best improve overall QoE.
\item To solve the mixed integer NP-hard problem, we propose algorithms to find promising solutions. Particularly, we propose a novel heuristic algorithm which can solve the NP-hard problem effectively with very low complexity.
\item We jointly consider the spatial and temporal domains. By investigating the impact of buffer length on FoV-driven 360-degree video transmissions, a novel hierarchical buffer updating strategy is proposed to ensure a robust buffer size with relatively high utility.
\item We show via simulations with 360-degree videos that the proposed algorithms yield significant QoE improvement over existing counterparts. The amount of performance enhancement is more pronounced when the network is crowded.
\end{enumerate}

\begin{table}[t]
\caption{Symbols and Notations} \vspace{-0.2cm}
\begin{center}
%\begin{tabular}{|c|c|}
\begin{tabular}{| >{\bfseries }l  p{7 cm } |}
\hline\hline
  $n$ & User index\\
  \hline
  $N$ & Total number of users\\
  \hline
  $U_{n}$ & The Utility of user $n$\\
  \hline
  $A,B$ & Normalization coefficients of $U_{n}$\\
  \hline
  $j$ & Tile index\\
  \hline
  $J$ & Total number of tiles in one video\\
  \hline
  $A,B$ & Normalization coefficients of $U_{n}$\\
  \hline
  $m$ & Tile representation level index\\
  \hline
  $M$ & Total Number of tile representation\\
  \hline
  $D_{m}$ & Video rate of tile representation $m$\\
  \hline
  $\theta_{0}$ & The prediction filed of view on sphere\\
  \hline
  $\varphi$ & Azimuth angle in spherical coordinate\\
  \hline
  $\theta$ & Polar angle in spherical coordinate\\
  \hline
  $\rho$ & Guarantee probability of FoV prediction \\
  \hline
  $y$ & Probable FoV index\\
  \hline
  $Y_{n}$ & Total number of probable FoV for user $n$\\
  \hline
  $P_{y}^{(n)}$ & The probability of probable FoV $n$ for user $n$  \\
  \hline
  $D_{n,j}$ & Video rate of tile $j$ on user $n$\\
  \hline
  $m$ & Tile representation index\\
  \hline
  $M$ & Total number of tile representation in server\\
  \hline
  $d_{n}$ & Total transmission rate of user $n$ through LTE and WLAN\\
  \hline
  $d_{n}^{LTE}$ & Transmission rate of user $n$ through LTE channel\\
  \hline
  $d_{n,i}^{wifi}$ & Transmission rate of user $n$ on AP $i$\\
  \hline
  $r_{n}^{LTE}$ & Achievable rate of user $n$ through LTE channel\\
  \hline
  $r_{n,i}^{wifi}$ & Achievable rate of user $n$ when connected to AP $i$\\
  \hline
  $W_{n,j}$ & Saliecny weight of tile $j$ on user $n$\\
  \hline
  $C_{n,j}^{m}$ & Cost of tile $j$ on user $n$ when $m$-th representation is selected\\
  \hline
  $\tilde{U}_{n,j}^{m}$ & Improvement utility of tile $j$ on user $n$ when $m$-th representation is selected\\
  \hline
  $\nu_{n,j}^{m}$ & Utility over cost of tile $j$ on user $n$ when $m$-th representation is selected\\
  \hline
  $B_{1}$ & Buffer threshold length\\
  \hline
  $B_{2}$ & Buffer maximum length\\
  \hline
  $B_{c}$ & Current buffer length\\
  \hline
  $\mu$ & Coefficient of QoE metric\\
  \hline
  $l$ & Coefficient of buffer strategy\\
  \hline
  $\sigma$ & Coefficient of penalty function\\
\hline \hline
\end{tabular}
\end{center}
 \label{tab:notation}
 \vspace{-0.7cm}
 \end{table}

The rest of the paper is organized as follows: Section II introduces tile-based 360-degree VR video and the system framework. Section III presents the problem formulation. Section IV tackles the problem and proposes effective algorithms to solve it. Section V addresses the buffer management problem in time-varying network with novel buffer strategy. Experiment results and performance evaluations are shown in Section VI, followed by conclusions and future works in Section VII.

\emph{Notations}: The symbols and notations used in this paper are summarized in Table I.

\section{Preliminary and System Model}\label{Sec_Preliminary and System Model}
\subsection{Tile-Based 360-degree VR Video}

\begin{figure}
\centering
\subfigure[]{
\includegraphics[width=3 in]{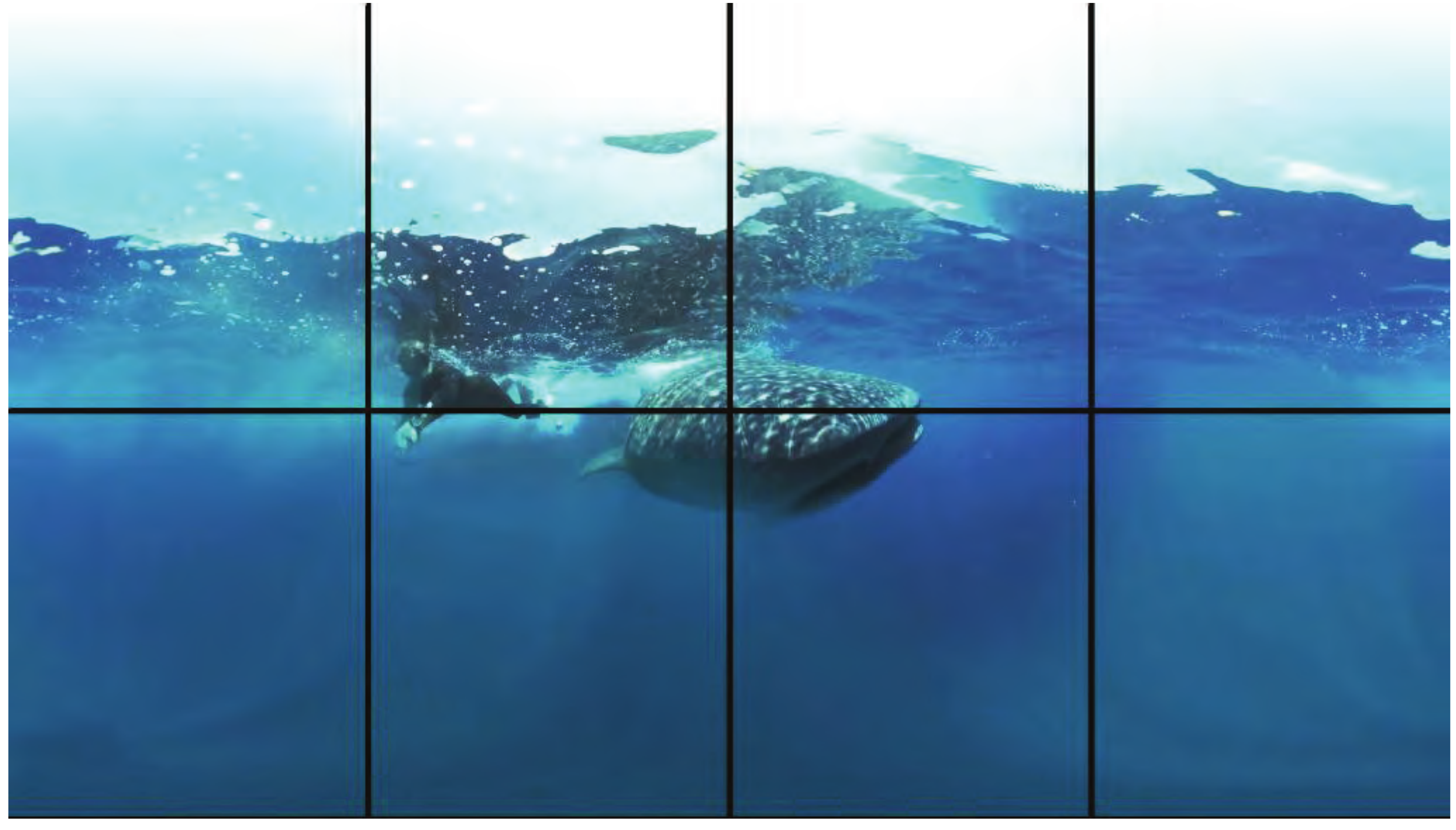}
}
\subfigure[]{
\includegraphics[width=3 in]{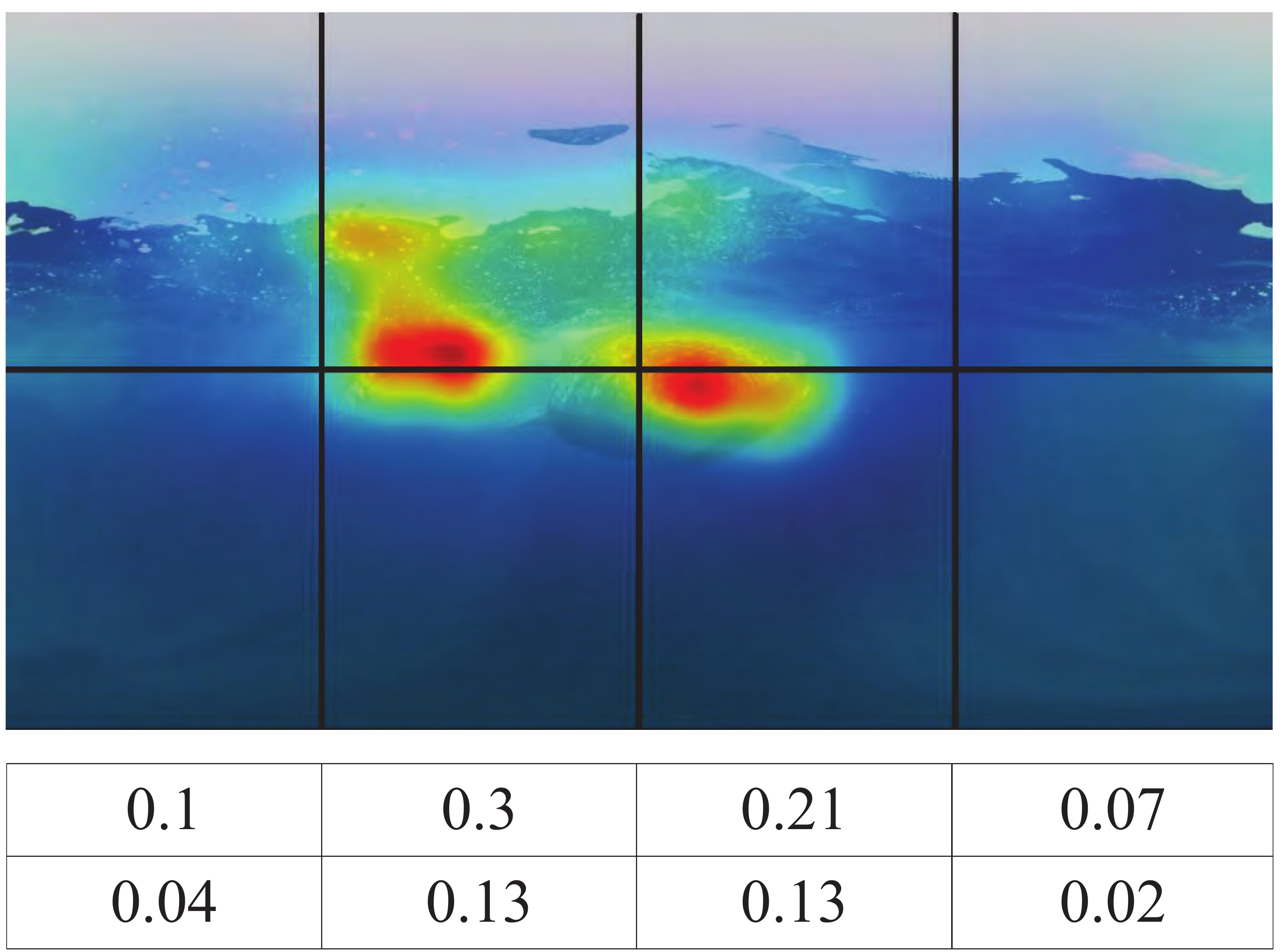}
}
\caption{(a) A 360-degree video is partitioned into 8 titles. (b) The saliency weights associated with each title. }\label{HW_figure_VR_video}
\end{figure}

To make a tiled 360-degree VR video, one can resort to either multiple source camera setup or partitioning of  a single video into multiple frames of smaller resolution. An example of a tile-based 360-degree VR video is shown in Figure 1(a), where the video is cut into 8 tiles, By specifying spatial relationship description in the file, the server can assign different networking resources to those tiles. In this paper, we jointly consider the saliency in the videos, FoV of users and various channel status to allocate the resources.

Saliency can be used in 360-degree videos to further improve the QoE \cite{S1,S2}. Compared with saliency in traditional videos, saliency in 360-degree videos also takes into account the position and projection \cite{saliency1,saliency2,saliency3}, so that a saliency map can highlight the regions where most people will look at. An example of the saliency map is shown in Figure 1(b). On one hand, users have a higher chance to look at those parts with higher saliency. On the other hand, when users looking at the higher-saliency portion of video, they will look more carefully on those parts, where distortions are easily perceived as more annoying and as such will receive a lower subjective quality score \cite{saliency4}. Therefore, we need to allocate more rates on tiles with higher saliency weights.

However, a user can only view a part of the whole 360-degree video at a certain time, i.e., the FoV. A 360-degree video may have several high-saliency weighted tiles which are not in the FoV. Besides, some users may have their own interests to look at videos. During the transmission, those tiles out of the FoV will be meaningless to have higher rates. To best utilize the bandwidth, we should consider the saliency in the FoV to improve the QoE.

\subsection{Utility Model}
We use the QoE as the performance metric in our paper, since it has been widely  used for video networking quality evaluation~\cite{QoE4,QoE0,QoE}. It is known that the user experience is not linearly proportional to the video rate, since it saturates at higher rates~\cite{QoE2,QoE3}. Therefore, the utility according to the rate can be defined as~\cite{QoE1}.

\begin{equation}
U(D)=A\operatorname{log}B\frac{D}{D_M},
\end{equation}
where $U$ denotes the utility of the video, $D$ denotes video rate; video rate $D\in\{D_1,\cdots,D_M\}$ belongs to one of the $M$ predefined DASH-VR rate representations\cite{MPD,representation} with $D_M$ being the maximum rate the server can provide.

When it comes to 360-degree video, as we illustrated, the utility is only meaningful in the FoV. Besides, the final utility in the FoV is not only related to the sum of utilities  of all tiles. More specifically, it  has been pointed out when the difference of rates between neighboring tiles is too large, the user can  be disturbed by the lowest rate in the FoV \cite{arxiv}. The overall QoE will be mainly affected by that tile with bad quality even though the rest of the tiles are of good qualities. Based on such definition, combing FoV and saliency weighting, we define the video quality of a user $n$ as:

\begin{equation}
\sum_{j\in\text{FoV}}{U_{n,j}*W_{n,j}+\mu\text{min}_{j\in\text{FoV}}(U(D_{n,j})),
}
\end{equation}
where $\mu\geq0$ is a tradeoff coefficient, $j$ is the index of tiles in each video. $D_{n,j}$ denotes the rate of tile $j$ in user $n$. The first part is the traditional utility used in 360-degree video, which sums the individual utilities of all tiles in the FoV. $W_{n,j}$ is the saliency weight for the tile $j$ in user $n$, which denotes the importance of visual perception the tile has. With the method in \cite{saliency1,saliency2}, we can derive the saliency score of each pixel in the 2D-screen for a 360-degree video. By giving normalized score according to the saliency results and calculating the sum of every pixel in each tile, we can derive the saliency weight of each tile as shown in Figure 1(b). The second part is used to make sure that the differences of rates in user's view are not too large. Thus, users will have a pleasant experience as we shown in the experimental results. The tradeoff coefficient $\mu$ should be content dependent and is empirically determined in this paper.

\subsection{The system of Tile-based 360-degree Video Transmission in Multi-RAT Network}
Thanks to the easy deployment of Wi-Fi APs, it is common that users have access to several APs in the coverage of a LTE station. To best explore the resource and utilize Wi-Fi APs around users, we consider a heterogeneous network to unicast multiple 360-degree videos to multi-users. The schemes and algorithms we proposed can also be extended to multi-casting and broadcasting scenarios by utilizing grouping methods. Figure 2 illustrates the detailed transmission of  tile-based 360-degree VR videos. At the server side, the raw 360-degree videos will be divided into tiles after projections. The encoder will generate different kinds of representations for each tile. Saliency detection is also carried out at the server side to get the saliency weight of each tile. All the tiles of 360-degree videos will be transmitted through a heterogeneous LTE/WLAN multi-radio network. In this network, all users can gain access to the LTE base-station (BS) and one of the Wi-Fi APs simultaneously. This scenario can be realized by LTE-WLAN aggregation technology \cite{LWA1,LWA2} or SDN-based LTE-WLAN multi-radio networks\cite{SDN2}. In this paper, our proposed scheme and algorithms are focused on APP-layer design, the architecture and the physical layer to support such centralized heterogeneous system is out of the scope of this paper. At the client side, all tiles will be combined together after decoding. Then rendering is used to help the 360-degree videos to present to users. Buffer model is also applied to cope with the problem when the channel of network is varying. Besides, a feedback link is used to derive the achievable rate based on the channel state information of each user \cite{feedback1,feedback2}, as well as the behavior of users (such as head-tracking results) for helping predict the FoVs. With all these information, the centralized heterogeneous network controller will help the video server to decide the rate allocation for each tile and the Wi-Fi AP association.

%the SDN controller is connected with the LTE packet data network gateway (PGW), and Wi-Fi APs are within the coverage of the LTE base station (BS).

\begin{figure*}
\centering
\includegraphics[width=6 in]{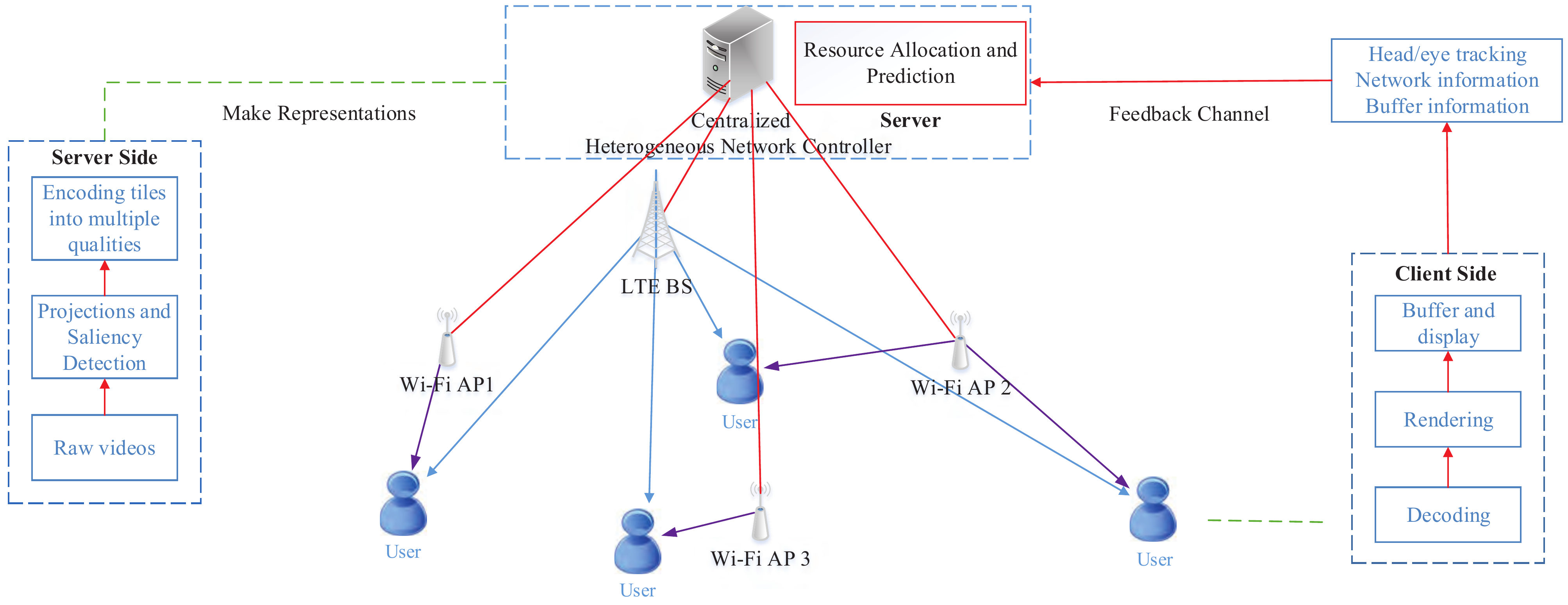}
\caption{The system model of the proposed tile-based 360-degree video streaming over a multi-RAT network.}\label{HW_figure_system_model}
\end{figure*}

\section{Resource Allocation in A Multi-RAT Network}

In the beginning of our transmission, users need initial frames to attract them to see the videos. Therefore, we allocate the rate on tiles based only on the derived saliency weights. With the transmission proceeding, users will have their own interests on the contents. Prediction is used and combined with the saliency weights to further save the bandwidth.  To avoid the user viewing a blank screen when the prediction algorithm is not reliable on some users, the server will transmit all the tiles at least in a very low representation. Our task is to decide the associations of users to Wi-Fi APs as well as the rate allocation for each tile through the heterogeneous network to maximize the total QoE. Thus, we formulate our FoV-driven transmission in a multi-RAT system with multi-users in this section.

\subsection{FoV Probability}
In this paper, we segment the 360-degree VR video into tiles and adopt equirectangular projection (ERP), which is widely used in 360-degree video data representation. There are techniques developed to estimate the FoV for a shorter duration ahead by analyzing users' viewing history and head movements (yaw and pitch as shown in Figure 3)~\cite{prediction,prediction1,prediction2}. However, the prediction may not be accurate, thus, if we use prediction with low accuracy, the total QoE of users will have a high probability of being extremely bad. Researches have also been done to give the predicted position $\theta_0$ on sphere view with the corresponding prediction accuracy $\gamma$~\cite{prediction,prediction1}.

Note that, throughout this paper, we assume that the FoV always contains the whole tiles exactly and if only part of a tile is in the FoV, it will be treated as a whole tile. Therefore, there are always integral number of tiles in the FoV. Importantly, when equirectangular projection is used, the number of tiles associated with the FoV is varying  depending on the position of FoV, e.g., the number  increases as the viewpoint deviates from the  equator. This can be resolved by rotating sphere geometry prior to projection,  however, which needs to prepare much more representations with an optimal set of rotation parameters. Therefore, without rotation with respect to the FoV, we need to map the predicted FoV on sphere (with $\phi$ and $\theta$ to describe the positions) to the tiles on 2D screen according to the omnidirectional projection relationship provided by MPEG as shown in Figure 3 \cite{projection}.

Combining the prediction angles, prediction accuracy and the projection relationship, we can derive the viewing probability of each probable FoV: $P=\mathcal{P}(\theta_0,\gamma,\phi,\theta)$. Note that based on the prediction, only several kinds of FoV have non-zero probability to be seen by users. We sort them with the probability in a descending order (indexed by $y$). They may contain different number of tiles based on their position, and here we only consider $Y_n$ kinds of predicted FoV which satisfy $\sum_{y}^{Y_n}{P}\geq\rho$, where $\rho$ is the guarantee probability. Then, the expected QoE of a specific user $n$ is:

\begin{equation}
\sum_{y}^{Y_n}{
	\left(\sum_{j\in\text{FoV}_y}{U(D_{n,j})}*W_{n,j}+\mu\min_{j\in\text{FoV}_y}~(U(D_{n,j}))\right)P_y
}
\end{equation}
where $P_{y}$ denotes the viewing probability of $FOV_{y}$ in a 2D screen. $\sum_{j\in\text{FoV}_y}{U(D_{n,j})}*W_{n,j}+\mu\min_{j\in\text{FoV}_y}~(U(D_{n,j})$ is the corresponding QoE for $FOV_{y}$ as illustrated in Eq. (2). We use the expected QoE as the objective function of the optimization throughout this paper.

\subsection{Problem Formulation}

Suppose there are one LTE, $I$ Wi-Fi APs (indexed by $i$), and $N$ users (indexed by $n$). Denote $r_n^\text{LTE}$ the maximum achievable rate of user $n$ based on its LTE channel quality, and $r_{n,i}^\text{wifi}$ the maximum achievable rate user $n$ can use when connected to Wi-Fi $i$.  Every 360-degree video is cut into $J$ tiles  (indexed by $j$). The video rate $D_{n,j}$ of each tile in each user can be chosen from $\{D_1,...,D_M \}$. $W$ denotes the weight of each tile based on saliency results. $P_y$ refers the viewing  probability of probable FoV. The optimization variables are LTE transmission rates $d_n^\text{LTE}, n=1,...,N$, Wi-Fi transmission rates $d_{n,i}^\text{wifi}, n=1,...,N, i=1,...,I$, the video rates $D_{n,j},n=1,...,N, j=1,...,J$ for each tile of each user, and users' associations on Wi-Fi AP. To maximize the expected QoE of all users according to our QoE metric defined in Eq. (3), the tile-based 360-degree video transmission problem formulation can be written as follows:

\begin{align}
\text{OPT}&\text{-1}:\nonumber\\
\max&\sum_{n}^{N}{\!\left(\!
\sum_{y}^{Y_n}\!
\left(\!
\sum_{j\in\text{FoV}_y}\!{U(D_{n,j})\!W_{n,j}}\!+\!\mu\!\min_{j\in\text{FoV}_y}\!{U(D_{n,j})}
\right)\!P_y\!
\right)
},\nonumber\\
{\mathrm{s.t.}}~&\sum_{j}{D_{n,j}}\leq d_n,~\forall n,\\
&~d_n=d_n^{\text{LTE}}+\sum_{i}{d_{n,i}^{\text{wifi}}},\\
&\sum_{n}{\frac{d_n^{\text{LTE}}}{r_n^{\text{LTE}}}}\leq 1,\\
&\sum_{n}{\frac{d_{n,i}^{\text{wifi}}}{r_{n,i}^{\text{wifi}}}}\leq 1,~\forall i,\\
&~\operatorname{card}(\left[d_{n,1}^{\text{wifi}},\cdots,d_{n,I}^{\text{wifi}}\right])=1,\\
&~D_{n,j}\in\{D_1,\cdots,D_M\}.
\end{align}

Eq.~(4) implies that for a certain user, the sum of the video rates of each tile $D_{n,j}$ should not exceed the total transmission rate $d_{n}$ allocated to it. Video rates of each tile can only be chosen from the representations in the server as specified in Eq.~(9).  The transmission rate can be aggregated from LTE and WLAN networks as illustrated in Section II.B. Eq.~(6) models the competition among all users for the limited bandwidth of LTE network: the sum of all transmission rates $d_n^{\text{LTE}}$ normalized by $r_n^{\text{LTE}}$ is upper bounded by $1$. The competition for the Wi-Fi APs is specified by the constraints in Eq.~(7). The constraint in Eq.~(8) means only one element of $\{d_{n,1}^{\text{wifi}},\cdots,d_{n,I}^{\text{wifi}}\}$  is nonzero, which enforces that each user $n$ can only be connected to a single Wi-Fi AP.

\section{Resource Allocation Algorithms}
The method proposed in \cite{Bell} is effective to solve the 360-degree video transmission with a single user. However, it is not applicable to such scenario with multi-user and multi-RAT taken into consideration, since the bandwidth/rate allocated to each user is unknown. Because the FoV and the content viewed by users can  influence the rate allocation, we cannot consider the rate for a user in a heterogeneous network and the rate for each tile separately. A systematic algorithm needs to decide them simultaneously. Besides, as seen from Eq.~(5), the transmission rate $d$ is jointly contributed by the Wi-Fi and LTE rates for each user, with only one Wi-Fi being chosen for each user. As a result, we can calculate the total QoE for \textit{any} Wi-Fi association so as to check all possible associations to find the optimal solution which also satisfies all the constraints. It is a mixed-integer NP-hard problem which similar to the well-known traveling salesman problem (TSP)\cite{TSP}. In the following, we consider to relax the discrete search space of OPT-1: $D_{n,j}\in\{D_1,\cdots,D_m\}$ to continuous search space $D_1\leq D_{n,j}\leq D_m$, so as to make the problem computationally tractable. Note that, after the relaxation, if we fix the Wi-Fi AP connection for each user, the problem becomes convex and thus can be solved by the convex optimization methods. An exhaustive search algorithm can be used to test all the possible Wi-Fi associations and perform the rate optimization for each association. Although the exhaustive search can guarantee to locate the optimal solution of AP association, it has a complexity of $\mathcal{O}(I^N)$.

\begin{figure*}
\centering
\includegraphics[width=5.5 in]{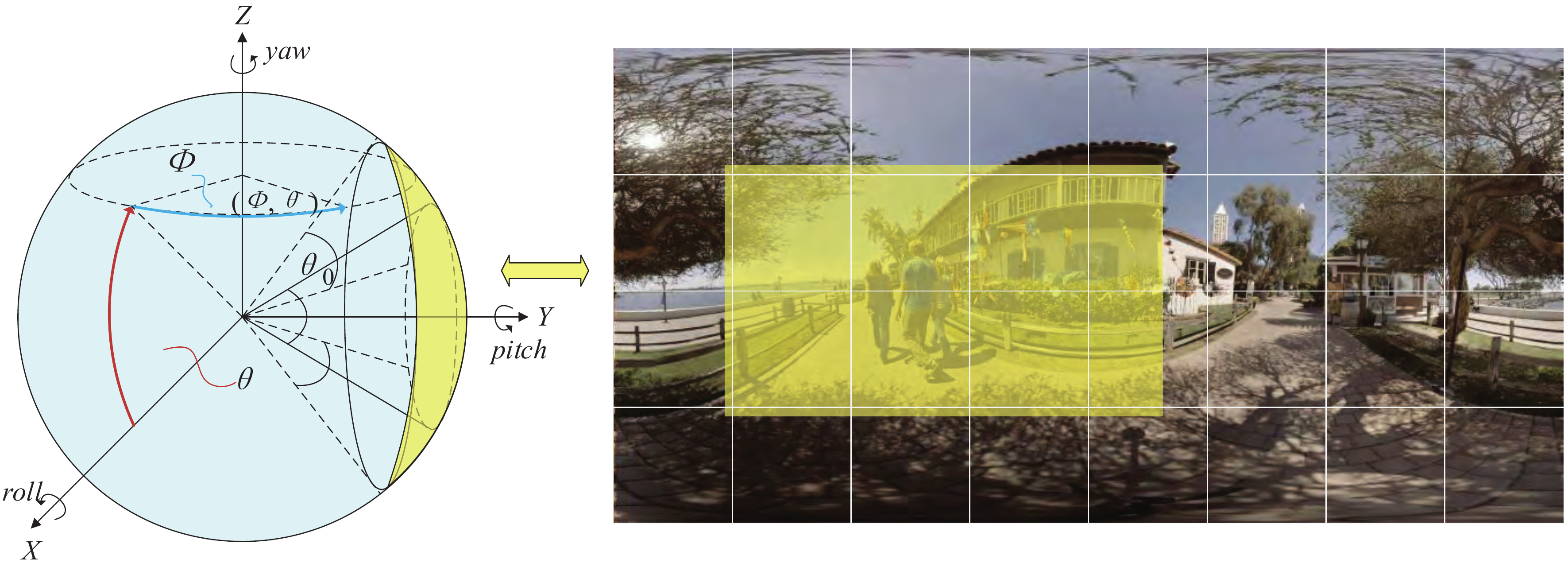}
\caption{FoV projection and rotating sphere geometry.}\label{HW_figure_Fov_projection}
\end{figure*}

\subsection{Greedy Algorithm}

\begin{algorithm}
%\begin{algorithm*}
\caption{Greedy Algorithm}
\label{Alg_Greedy Algorithm}
\begin{algorithmic}[1]
\renewcommand{\algorithmicrequire}{\textbf{Variable definition:}}
\REQUIRE {\quad \\
$\mathcal{Q}_B$: the set of fixed users and corresponding association;\\
$\mathcal{Q}_C$:the set of users have not been placed;\\
$(n,i)$: the user $n$ is connected to Wi-Fi AP $i$};

\renewcommand{\algorithmicrequire}{\textbf{Initial:}}
\REQUIRE {
	$\mathcal{Q}_B=\Phi,\mathcal{Q}_C=\{1,2,\cdots,N\}$
};

\WHILE{$\mathcal{Q}_C!=\Phi$}
\STATE{$sum=0;j=0;t=0$;}
\FOR{$n=1,n\leq N,n++$}
\IF{$n\in\mathcal{Q}_C$}
\FOR{$i=1,i\leq I,i++$}
\STATE {solve the OPT-1 when user $n$ is fixed to Wi-Fi AP $i$;\\
$temp=\max{U_{\mathcal{Q}_{B}+\{(n,i)\}}}$,\\
$D^*\gets \text{arg}\max{U_{\mathcal{Q}_{B}+\{(n,i)\}}}$;}
\IF{$temp>sum$}
\STATE{$sum=temp;j=n,t=i;$}
\ENDIF
\ENDFOR
\ENDIF
\ENDFOR
\STATE \textbf{update} $\mathcal{Q}_{B}\gets \mathcal{Q}_{B}+\{(j,t)\};\mathcal{Q}_{C}\gets \mathcal{Q}_{B}-\{j\}$ and $D^*$;
\ENDWHILE
\STATE Round $D^*$ to appropriate representation.
\end{algorithmic}
\end{algorithm}
%\end{algorithm*}

To avoid the exhaustive search, we first propose a greedy algorithm to find a feasible solution, which  starts with the number of user is $0$ in the system.   It places each user into the system and fixes them on each Wi-Fi AP. Solve the OPT-1 for $N*I$ times to find one feasible solution that improves the objective function most. Then we repeat the search for $(N-1) *I$ times with the previous connection fixed. Repeat the iterations until all the users are assigned. In each iteration, it is a convex problem and can be solved by the convex optimization method.

Let $D^*$ be an optimal solution of the relaxed optimization of OPT-1. However, we have to discretize the solution as  $\{D_1,\cdots,D_m\}$. A simple algorithm is quantizing $D^*$ to the closet value $D^f$ that belongs to one of $\{D_1,\cdots,D_m\}$, and making sure that $D^f$ satisfies the constraint $\sum_{j}{D_{n,j}^f}\leq \sum_{j}{D_{n,j}^*}$, otherwise, lower the level of $D^f$. Note that some bandwidth will be wasted after the quantization in some cases. If we want to utilize the wasted bandwidth, we need to further consider which user can utilize the bandwidth effectively based on their achievable rates on each network. We can allocate rates to the tile which can improve the objective function most. However, through simulations, we found that since our method is a centralized method which fully utilizes the resources and we adopt a large number of representations, the feasible solution obtained by quantizing method has already been very promising in most cases. This further step can only achieve a small promotion with increasing the complexity. Thus, through this paper, we will just quantize the result without further utilizing the small wasted bandwidth. The greedy algorithm is illustrated in Algorithm~\ref{Alg_Greedy Algorithm}.

\subsection{Heuristic Algorithm with Penalty Function}

\begin{algorithm}[t]
%\begin{algorithm*}
\caption{Heuristic Algorithm with Penalty Function}
\label{Alg_Heuristic Algorithm}
\begin{algorithmic}[1]
\STATE Convert OPT-1 to OPT-2 with penalty function;
\STATE Solve OPT-2 and get $d_{n,i}^{\text{wifi}}$;
\STATE Find the $K$ users (indexed by $k$) connected to more than a single AP;
\FOR{$k=1,n\leq K,k++$}
\STATE Use greedy approach only on these users while fixed others and update $D^*$;
\ENDFOR
\STATE Round $D^*$ to appropriate representation.
\end{algorithmic}
\end{algorithm}
%\end{algorithm*}

However, the greedy approach still has a complexity of $\mathcal{O}(N^2I)$, which is not effective as will be evidenced in the simulations. To further reduce the algorithmic complexity and improve performance, we propose a provably near-optimal solution by  introducing a penalty function into the problem and relaxing the OPT-1 into a convex problem. To be specific, the penalty function is a regularization term, which is the square root of the $\ell_1$ norm of vectors $\Vert[d_{n,1}^{\text{wifi}},\cdots,d_{n,I}^{\text{wifi}}]\Vert_1$. Thus, the problem can be written as:

\begin{align}
\text{OPT}&\text{-2}:\nonumber \\
\max &\sum_{n}^{N}\left(\sum_{y}^{Y_n}(\mu\!\!\min_{j\in\text{FoV}_y}{U(D_{n,j})}+\sum_{j\in\text{FoV}_y}\!{U(D_{n,j})W_{n,j}}
)P_y
\right)\nonumber \\
&-\sigma\sqrt{
\sum_{n}\left(\sum_{i}{d_{n,i}^{\text{wifi}}}\right)^2
}\nonumber\\
{\mathrm{s.t.}}~&\sum_{j}{D_{n,j}}\leq d_n,~\forall n,\\
&\sum_{n}{\frac{d_n^{\text{LTE}}}{r_n^{\text{LTE}}}}\leq 1,\\
&\sum_{n}{\frac{d_{n,i}^{\text{wifi}}}{r_{n,i}^{\text{wifi}}}}\leq 1,~\forall i,\\
&~d\geq 0,\\
&~D_1\leq D_{n,j}\leq D_m.
\end{align}
where $\sigma$ is the coefficient of the penalty function and we empirically set it as $0.1$ through this paper. The cardinality $\operatorname{card}([d_{n,1}^{\text{wifi}},\cdots,d_{n,I}^{\text{wifi}}])=1$ constraint for each user $n$ is  relaxed as an $\ell_1$ norm constraint, i.e., $\Vert[d_{n,1}^{\text{wifi}},\cdots,d_{n,I}^{\text{wifi}}]\Vert_1=\sum_{i}{d_{n,i}^{\text{wifi}}}\leq1$. Note that $\ell_1$ norm constraint promotes the sparsity of the vector $[d_{n,1}^\text{wifi},\cdots,d_{n,I}^\text{wifi}]$, and forces many $d_{n,i}^{\text{wifi}}$ to be zero. Instead of imposing a number of strict constraints, we penalize the Euclidean norm of the $\ell_1$ norm, which can collectively force $[d_{n,1}^\text{wifi},\cdots,d_{n,I}^\text{wifi}]$ to be sparse. In many cases, only one nonzero element that gives an assignment of user $n$ to a specific Wi-Fi $i$  is obtained. Based on this new formulation, we can get the video rates $D_{n,j}$ on each tile of each user if the problem is convex.

Thus, we theoretically analyze the OPT-2 and find the objective function is convex, the set of the constraints are convex and the Slater condition is satisfied. It is a convex optimization problem which can be effectively solved by existing convex optimization methods~\cite{convex1,convex2}.

Notice that the solution can potentially make users connect to more than one AP after the relaxation of the constraints. Although some users are still assigned to more than a single AP, we can find that due to the penalty function, only one element of $[d_{n,1}^\text{wifi},\cdots,d_{n,I}^\text{wifi}]$ is large enough and the rest  are relatively small. We can identify users who are still connected to more than one AP with $d_{n,i}^{\text{wifi}}$ larger than a small pre-set threshold. After that, we can use search methods on users with more than one assigned APs while keeping the associations  of other users fixed to find a sub-optimal solution. Then we apply the quantization method, same as that of the Algorithm 1, to discretize the solution as $\{D_1,\cdots,D_m\}$. The procedure is summarized in Algorithm~\ref{Alg_Heuristic Algorithm}.

From Algorithm~\ref{Alg_Heuristic Algorithm}, we can find that the complexity of this heuristic algorithm is very low when compared with the greedy algorithm. According to our simulations, in a 15-user system, no more than $3$ users will be assigned to multiple APs after the first step. The algorithm can be included as a module of the multi-RAN controller, which decides the allocation and association for several frames based on the prediction.  For each iteration, convergence loop is continuously iterated from the previous point, rather than from an  initial point. Thus, the controller can respond to users' behaviors quickly with fast convergence.

\subsection{Decomposition Algorithm}

An algorithm that decomposes the problem is proposed here to derive the solution as quickly as Algorithm~2 without relaxing the discrete strategy space. More specifically, the problem OPT-1 can be decomposed into two optimization problems, OPT-3 and OPT-4, as follows:

\begin{align}
\text{OPT}&\text{-3}:\nonumber \\
\max&\sum_n^N{
  \left(
  U_n(d_n)\sum_y^{Y_n}{(\sum_{j\in \text{FoV}_y}W_{n,j})P_y}
  \right)\quad\quad\quad\quad\quad\quad\quad\quad
}
\nonumber \\
{\mathrm{s.t.}}~&d_n=d_n^{\text{LTE}}+\sum_{i}{d_{n,i}^{\text{wifi}}},\\
&\sum_{n}{\frac{d_n^{\text{LTE}}}{r_n^{\text{LTE}}}}\leq 1,\\
&\sum_{n}{\frac{d_{n,i}^{\text{wifi}}}{r_{n,i}^{\text{wifi}}}}\leq 1,~\forall i,\\
&\operatorname{card}(\left[d_{n,1}^{\text{wifi}},\cdots,d_{n,I}^{\text{wifi}}\right])=1.
\end{align}

\begin{align}
\text{OPT}&\text{-4}:\nonumber \\
\max&\left(\sum_y^{Y_n}{\mu\min_{j\in \text{FoV}_y}~{U(D_{n,j})}}+ \sum_{j\in \text{FoV}_y}{U(D_{n,j})W_{n,j}}
\right)\quad\quad\quad\quad \nonumber \\
{\mathrm{s.t.}}~&\sum_j{D_{n,j}}\leq d_n,\\
&D_{n,j}\in\{D_1,\cdots,D_M\}.
\end{align}

OPT-3 combines the saliency weight and FoV probability with the channel state information of users, it is aiming to derive the Wi-Fi AP association and rate allocation for each user. OPT-3 can be effectively solved by applying penalty function-based method as Algorithm~\ref{Alg_Heuristic Algorithm}. The proposed problem OPT-4 is used to optimize the rate allocation on each tile for a certain user based on the results from OPT-3. We can solve the knapsack problem OPT-4 for each user with greedy approach similar to~\cite{Bell}. By sorting the utility over cost for each tile with probable representation $D_{n,j}\in\{D_1,\cdots,D_M\}$ and continuously updating the representations until all rates are consumed up, we can get the feasible solution $D_{n,j}$ for each tile of users. The cost function $C_{n,j}^{m}$ shows the cost to pay for choosing the $m$-th representation on tile $j$. After the first representation is selected, the algorithm can improve the representation, and it only needs to pay the difference between the allocated representation and the new representation. Therefore, it can be iterated quickly if we want to update the representation with more bandwidth/rates.

The cost is defined as:
\begin{equation}
C_{n,j}^{m}=\left\{\begin{matrix}D_{n,j}^{m}-D_{n,j}^{m-1}, &m\geq 2

\\ D_{n,j}^{m}, &m=1

\end{matrix}\right.
\end{equation}

The utility gain is defined as:
\begin{equation}
\tilde{U}(D_{n,j}^{m}))= \left\{\begin{matrix}U_{n,j}^{m}-U_{n,j}^{m-1}, &m\geqslant 2
\\
U_{n,j}^{m}, &m=1
\end{matrix}\right.
\end{equation}

The utility over cost denotes the utility gain when $m$-th representation is selected for tile $j$ of each user $n$ per cost, which is defined as:
\begin{equation}
\nu _{n,j}^{m} = \frac{\tilde{U}(D_{n,j}^{m}))W_{n,j}}{C_{n,j}^{m}/r_{n}},
\end{equation}
note that the utility over cost defined here is different from that in \cite{Bell}. It is because our system has multiple users, and different users have different extents of ability to utilize the bandwidth. The decomposition algorithm is summarized in Algorithm~\ref{Alg_Decomposition Algorithm}.

\begin{algorithm}[t]
%\begin{algorithm*}
\caption{Decomposition Algorithm}
\label{Alg_Decomposition Algorithm}
\begin{algorithmic}[1]
\STATE Solve OPT-3, get $d_{n}$ and Wi-Fi AP allocation;
\FOR {$n=1,n\leq N,n++$}
\STATE sort the utility over the cost of each tile $\nu _{n,j}^{m}$ and set $d_{\text{current}}=0$;
\WHILE{$d_{\text{current}}\leq d_n$}
\STATE \textbf{update} the representation level to each tile according to the utility over the cost continuously ;
\STATE \textbf{update} $d_{\text{current}}=d_{\text{current}}+d_{\text{consumed}}$;
\ENDWHILE
\ENDFOR
\end{algorithmic}
\end{algorithm}
%\end{algorithm*}

Although the algorithm decomposes the original problem OPT-1, it allows us to use discrete strategy space without relaxation. The results are also promising in some scenarios. What is more, the iteration in this algorithm can also converge quickly from current result and is useful for updating the buffer which will be discussed in next section.

Note that, if the server cannot utilize a feedback link to get the users' behavior information to obtain predict the FoV, all the three algorithms are still applicable by using broadcasting. By applying our algorithms without FoV prediction and FoV probability results, the server can broadcast each video to users just according to the saliency weight.

\section{Buffer Management Strategy}

Without a playback buffer, a user may see a frozen/blank screen and has to wait for the arrival of the subsequent video frames, if we cannot estimate the states accurately under a time-varying channel condition. This would  result in poor QoE. However, as illustrated in Section I, buffer management schemes for traditional videos are not appropriate for 360-degree videos due to the short-time prediction nature of FoV. To save bandwidth and attain high QoE, the FoV-driven schemes can only be applied in a short time from current view (about $2$s)~\cite{prediction}, while a robust buffer to avoid the re-buffering events requires a relatively long buffer length. Therefore, in this section, we propose a novel hierarchical buffer updating strategy, built upon our proposed resource allocation algorithms, to solve this problem for FoV-driven 360-degree videos.

\begin{figure}
\centering
\includegraphics[width=3.2 in]{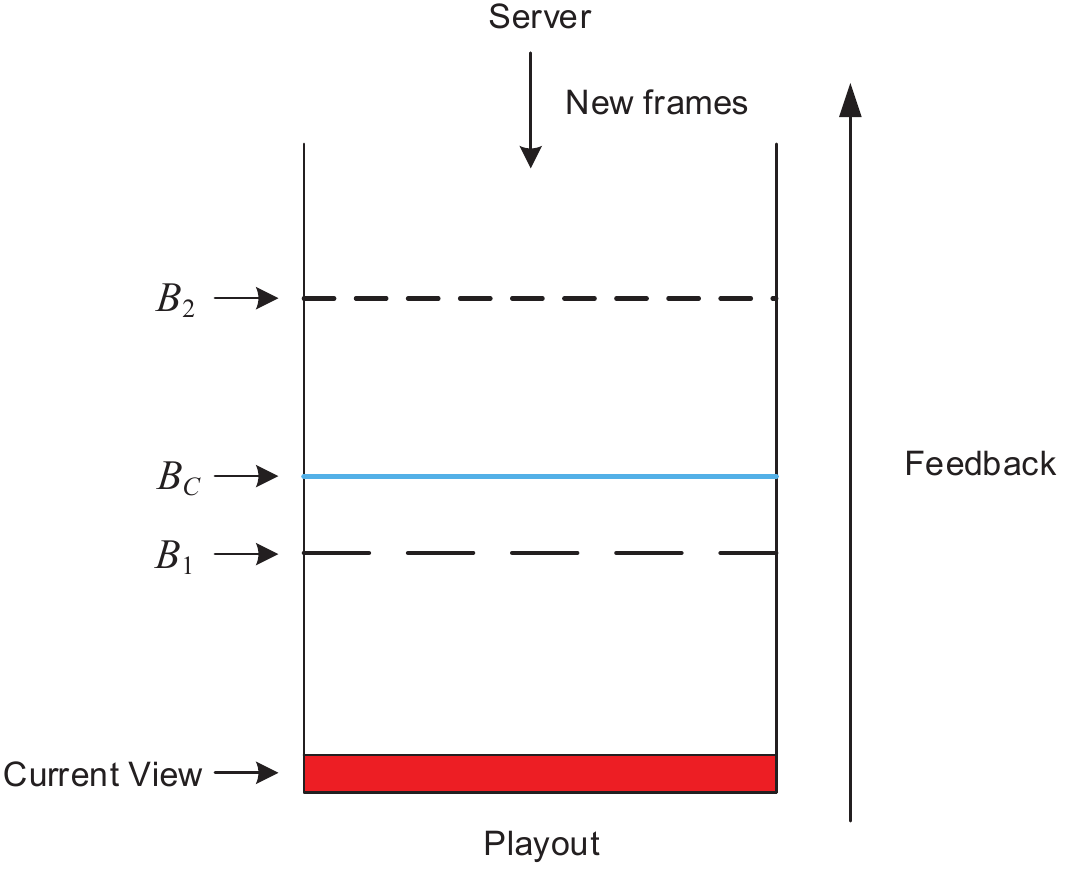}
\caption{A hierarchical updating buffer.}\label{HW_figure_Hierarchical}
\end{figure}

The hierarchical updating buffer is shown in Figure~\ref{HW_figure_Hierarchical}, where $B_1$ is the buffer length threshold that is set according to the lowest accuracy the server can accept. Therefore, the prediction-based scheme is acceptable only if the the buffer has less than $B_1$ frames. If not, it is hard to predict the behavior of users and the server needs to transmit all the tiles equally. $B_2$ is the maximum buffer length which is set to avoid the re-buffering events in time-varying networks.

We can adopt  transmission schemes with our proposed algorithms for frames during $[0, B_1]$ in the buffer, while only streaming the low representations equally for all tiles of the frames during $[B_1, B_2]$. However, when the frames based on the prediction are consumed up, users will see the low-quality frames even if the channel is good enough for higher quality frames. To solve this problem, we can update the qualities of tiles with low representation if we have more bandwidth for updating and the frames in the buffer are sufficient to avoid the re-buffering events. Thus, the rate/bandwidth we can allocate for subsequent optimization is decomposed into two parts, one is used for updating the arriving  tiles for playout to maximize the QoE, and the other   is used for storing more frames to minimize the probability of re-buffering events.

Instead of only using adaptive rate allocation for the subsequent optimization, we integrate our transmission algorithms with buffer-based scheme to achieve better QoE. Informally, we should make the rate selection more conservative and lower down the rate for updating the frames in the buffer when the buffer is at risk of underrunning. On the other hand, more aggressive rate selection and more rates on updating when the buffer is close to full. Thus, the rate we can allocate for the subsequent optimization is in proportion to the estimated throughput and current buffer size $d_{\text{subsequent}}=f(d_{\text{estimated}},B_c )$~\cite{buff4,buff5}. In this paper, we denote the proportional relationship as follows:
\begin{equation}
d_{\text{subsequent}}= l \tfrac{d_{\text{estimated}}}{B_{2}-B_{c}},
\end{equation}
where $l\geq 0$ is the proportional coefficient, which can be decided by sever as~\cite{buff4}.

Our hierarchical  buffer updating strategy for 360-degree VR videos can be summarized  as follows:

If the current buffer size satisfies $B_c<B_1$, the buffer is at a risk of running out. We use Algorithm~\ref{Alg_Heuristic Algorithm} or \ref{Alg_Decomposition Algorithm} to transmit frames during $[0, B_1]$. In this step, we only transmit the whole frames during $[B_c, B_1]$. For those frames during $[0, B_c]$, we only update those tiles that need a higher level representation based on the results and keep using other tiles already in the buffer, while the rates are equally allocated to tiles for the frames after $B_1$.

If the current buffer size satisfies $B_c>B_1$, the buffer is considered full enough. We will first use Algorithm~\ref{Alg_Heuristic Algorithm} or \ref{Alg_Decomposition Algorithm} to update the tiles during $[0, B_1]$. Then we transmit the tiles with equal rates for frames after $B_c$, until all estimated rates are consumed up.

This strategy can ensure enough tiles in the buffer and always update the upcoming playout tiles if there are enough rates for allocation. It makes a good trade-off between updating existing frames and downloading new frames. Besides, the maximum buffer length is a bit smaller than traditional video buffer to reduce the influence of bad prediction in a long-length buffer. In future work, we will investigate the optimal size of $B_{1}$ and $B_{2}$ to achieve better performance.

Since MPEG (Moving Picture Experts Group) has already supported the tiling scheme in the Media Presentation Description (MPD) file  \cite{MPD}, our buffer scheme can be easily applied to the system. By specifying spatial relationship description (SRD) in the MPD file, the server can transmit certain tiles with certain rates to users \cite{SRD}.

\section{Simulations}

In this section, the effectiveness and favorable performance of our proposed immersive media transmission scheme is validated via simulations. Overall, the achieved utilities of our proposed algorithms are much higher than other benchmark techniques. The amount of improvement in resource allocation is evaluated with bandwidth and number of users in the system. The performance of buffer strategy is evaluated through a time-varying network. We begin with describing the simulation setup used in the later evaluations.

\subsection{Simulation Setup}

To prove the efficiency of our scheme and show the generality, we use $18$ distinct 360-degree videos from MPEG-JVET (Joint Video Exploring Team) 360-degree VR video datasets and YouTube (as shown in Table. II) to evaluate our proposed scheme. All videos are used equirectangular projection and segmented into $4*8=32$ tiles. The server can provide $10$ different-bitrate  representations for each tile: $\{0.1, 0.2,\cdots, 0.9, 1\}$Mbps (indexed from $1$ to $10$) and the guaranteed probability $\rho$ is set as $0.95$. The frame rate is set as $30$fps, and $15$ frames are in one Group of Pictures (GoP). $B_{1}$ and $B_{2}$ are set as $2$s and $5$s respectively. The 360-degree videos are viewed by users through head-mounted displays (HMD), such as HTC Vives. The FoV is about 120*90 degrees. As we explained in Section III. A, we assume that the FoV contains the whole tiles exactly. The number of tiles in the FoV will be changed through the position due to the ERP we used. Thus, in our case the FoV consists $6$ tiles at least and $12$ tiles at most. Video contents are requested randomly by users, whose motions are recorded and converted to prediction results as method in \cite{prediction}. The transmission part is simulated in NS-3, IEEE 802.11n Wi-Fi APs and default LTE parameters are used in the module. The users are uniformly distributed around $5$ Wi-Fi APs within the coverage of LTE BS within $200$m. When more users are involved into the system, they will be set close to AP $1$ to simulate the congestion scenario. All users can access any one of the Wi-Fi APs and LTE at the same time and their achievable rates can be calculated based on positions and channel quality information. Total bandwidth can be changed from $10$MHz to $70$MHz.

\begin{table}[]
\centering
\caption{Videos used for evaluation}
\label{my-label}
\begin{tabular}{ll}
\hline
Name & Source Quality \\ \hline
AerialCity & 1080P \\
Balboa & 1080P/4K \\
BranCastle & 1080P/4K \\
Broadway & 1080P/4K \\
ChairliftRide & 4K \\
Diving with sharks & 4K \\
DrivingInCity & 1080P \\
DrivingInCountry & 1080P \\
Gaslamp & 1080P/4K \\
Harbor & 4K \\
KiteFlite & 4K \\
Landing & 4K \\
Polevault & 1080P \\
SkateboardInLot & 4K \\
SkateboardTrick & 4K \\
Surrounded by Wild Elephants & 4K \\
Train & 4K \\
Trolley & 4K \\ \hline
\end{tabular}
\end{table}

To prove the superior performance, we compare our scheme \textbf{(Centralized + Probable FoV + Algorithm~\ref{Alg_Greedy Algorithm}, Centralized + Probable FoV + Algorithm~\ref{Alg_Heuristic Algorithm} and Centralized + Probable FoV + Algorithm~\ref{Alg_Decomposition Algorithm})} with the following competing schemes:

\begin{enumerate}

\item \textbf{Centralized + One FoV + Algorithm~\ref{Alg_Heuristic Algorithm}:} It utilizes the heterogeneous network we mentioned in Section II, which can centralized control the resource to transmit the videos. However, only the most probable FoV is taken into consideration  without prediction accuracy (FoV probability results).

\item \textbf{Decentralized + Probable FoV:} It considers all the factors as our proposed method except it optimizes the LTE and WLAN resource separately in a decentralized network.

\item \textbf{Centralized  + Probable FoV + Exhaustive Search:} It utilizes our heterogeneous network to transmit the videos. However, it  tests all possible Wi-Fi AP associations and optimizes resource allocation for each.

\item \textbf{Centralized  + Probable FoV + Equal Rate Allocation:} It utilizes our heterogeneous network to transmit the videos. However, it allocates rates equally in the FoV and  ignores the saliency contribution.

\item \textbf{Short Buffer-Based Strategy:} It utilizes our Algorithm~\ref{Alg_Heuristic Algorithm} in streaming, but the buffer is only $2$s length. Therefore, the prediction accuracy of FoV is acceptable for all frames in the buffer.

\item \textbf{Long Buffer-Based Strategy:} It is similar to short buffer-based scheme, whereas the buffer size is $5$s. The prediction accuracy of FoV decreases with the increase of contents in buffer. For all the frames with prediction accuracy is $0$, the server will transmit equal rates to all tiles.
\end{enumerate}

\subsection{Simulation Results}

\begin{figure}
\centering
\subfigure[]{
\includegraphics[width=3 in]{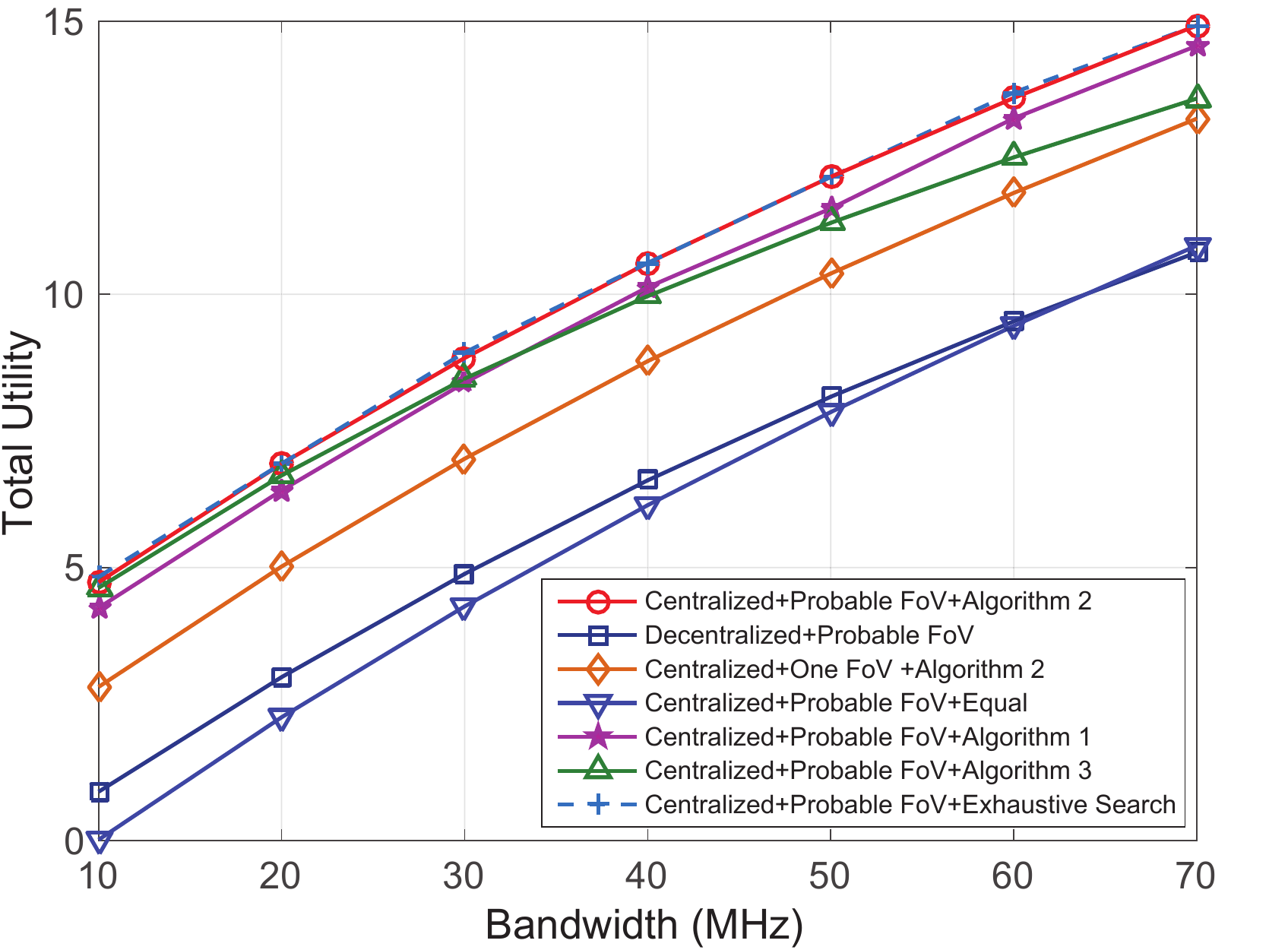}
}
\subfigure[]{
\includegraphics[width=3 in]{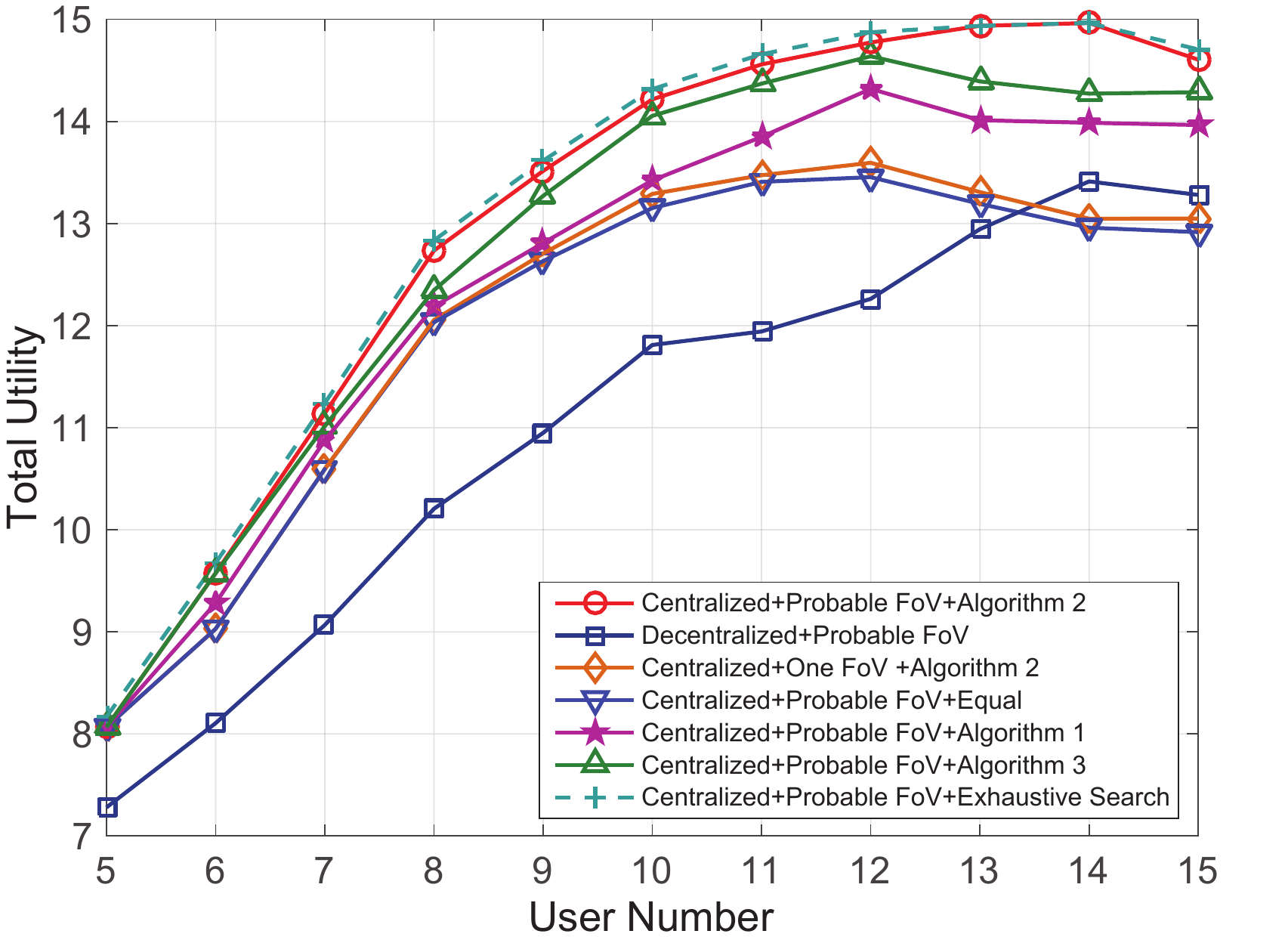}
}
\caption{The performance of different methods. (a) Utility with bandwidth. (b) Utility with number of users. }\label{HW_figure_performance}
\end{figure}

Figure 5 illustrates the utility of all methods with respect to the available bandwidth and number of users. We can clearly see that our proposed scheme performs much better than others, especially the performance of our scheme with Algorithm~\ref{Alg_Heuristic Algorithm} is very close to that of the exhaustive method. Our scheme with Algorithm 3 performs well when the bandwidth is small as shown in Figure 5(a). However, it separates the problem and tries to fetch the bandwidth first for users. As a result, the performance is not so promising when there is enough bandwidth to improve the quality in the FoV. Besides, our scheme with Algorithm~\ref{Alg_Greedy Algorithm} also performs well until the number of users increases to $13$ (see  Figure 5(b)), where congestion starts to occur on Wi-Fi AP $1$.

It is worthwhile to point out that  the total utility (QoE) begins to drop because the server wants to ensure every user get at least a lowest representation. Users cannot get enough rates on high saliency parts. Consequently, the total QoE decreases.  Our scheme with heuristic algorithm drops (until the number of users increases to $15$) later than others, since it can consider the saliency, FoV prediction and users' channel states collectively. As expected, decentralized method performs badly because it utilizes the two networks separately. When there is a congestion in one of the network, the server cannot allocate the resource effectively.

The simulation results also reveal that saliency is a significant factor to influence the QoE when users cannot get enough rates. When there are $15$ users in the system, the utility of equal rate allocation method is even worse than decentralized method. In such scenario, each user can only get small rates due to the congestion, the equal allocation method wastes the limited resource on those lower saliency parts, resulting in a poor utility. On the other hand, when there are less users or more bandwidth, equal rate allocation becomes better than decentralized method.

\begin{figure}
\centering
\subfigure[]{
\includegraphics[width=3.2 in]{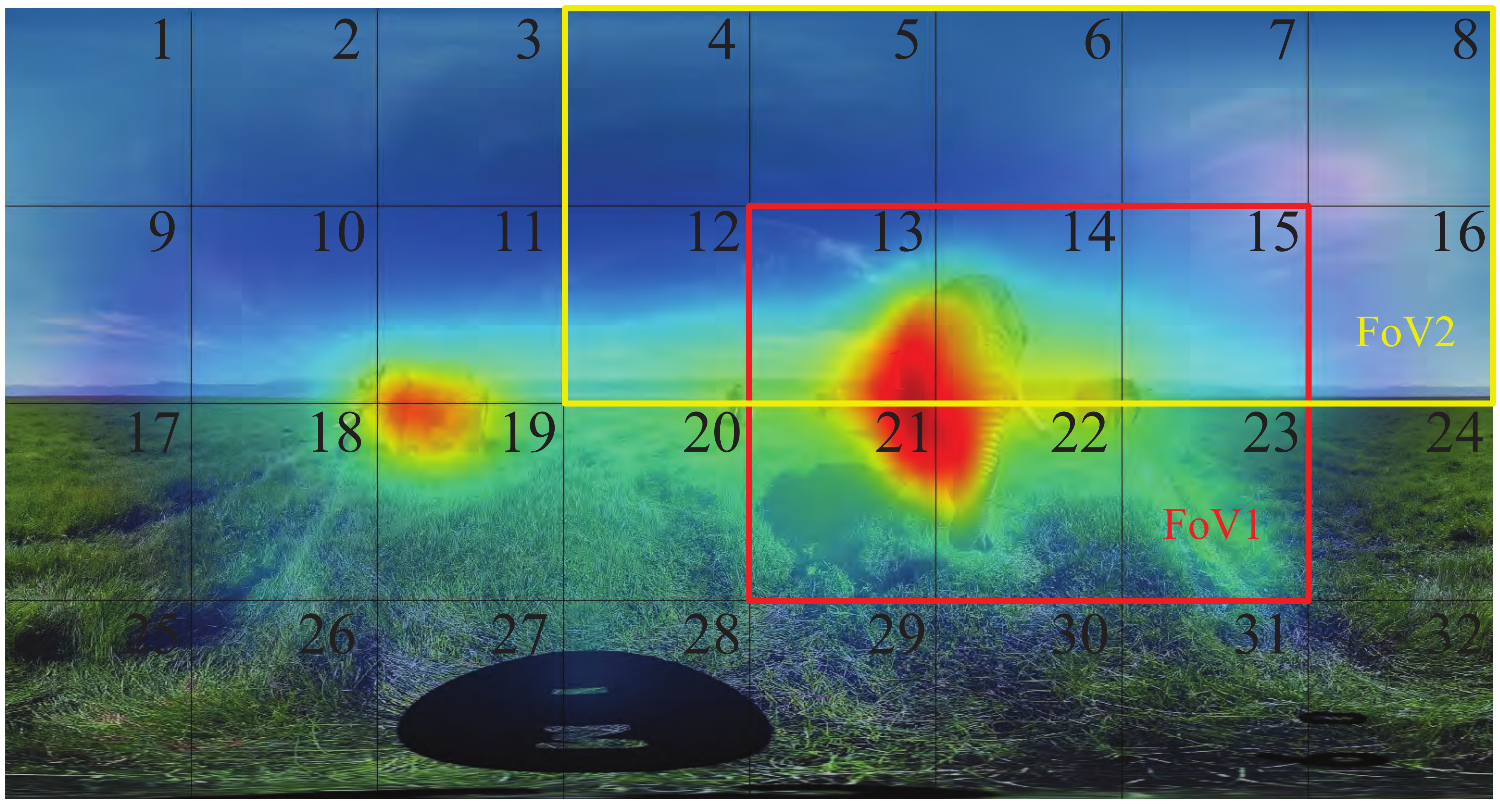}
}
\subfigure[]{
\includegraphics[width=3.2 in]{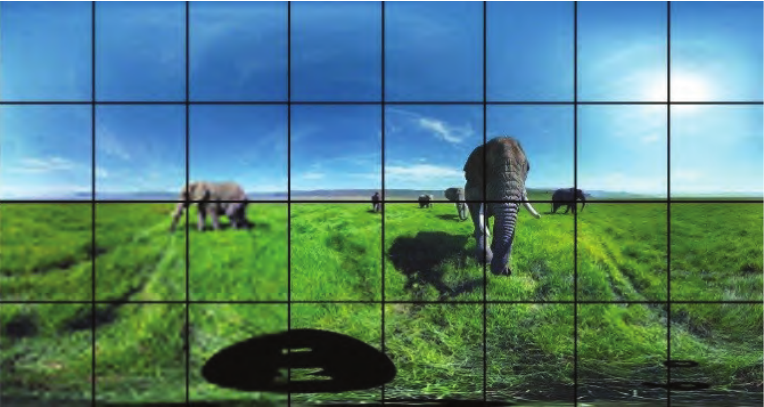}
}
\caption{(a) The saliency and FoV example. (b) The performance of 360-degree video.}\label{HW_figure_video_performance}
\end{figure}

\begin{table}[]
\centering
\caption{The representation level of each tile in Figure 6}
\label{my-label}
\begin{tabular}{llllllll}
\hline
1 & 1 & 1 & 2 & 2 & 2 & 2 & 2 \\
1 & 1 & 1 & 3 & 9 & 8 & 7 & 2 \\
1 & 1 & 1 & 3 & 9 & 8 & 6 & 4 \\
1 & 1 & 1 & 3 & 4 & 5 & 5 & 4 \\ \hline
\end{tabular}
\end{table}

\textbf{Impact of Saliency and FoV:} We demonstrate the video performance with our scheme (Heterogeneous + Probable FoV + Algorithm~\ref{Alg_Heuristic Algorithm}) in Figure 6. Figure 6(a) shows the saliency map and two exemplar  FoVs among all probable FoVs from the prediction of one user. Figure 6(b) demonstrates the video quality of all tiles. It can be seen that FoV 1 has the highest quality due to the high probability and high saliency weight.

Although FoV 2 has a similar  viewing probability as FoV~1, we can find FoV~2 contains more tiles since it is close to the pole on the sphere. Besides, the levels of representations allocated to  FoV~2 are smaller than on FoV~1 as shown in Table III. It is because that the saliency weight in FoV~1 is much larger than that in FoV~2. By contrast, even tile 6 has a low saliency weight, due to the QoE metric used, the rate difference in the FoV will not be big. Since users are more likely to be attracted by the high saliency object (e.g., the elephant) in the FoV, it is hard to detect the visual discrepancy  with small rate difference as shown in Figure 6(b). There is another elephant we can see on the left of the 2D screen. Despite the high saliency on that, it is not in the predicted FoV, resulting in a lowest rate.  What is interesting, tiles 4-8's rates are much lower than others'. However, most users can still enjoy the video even they are viewing FoV~2, it is hard to detect the visual discrepancy since they will not put more attention on the low saliency object (e.g. the sky).

\begin{figure}
\centering
\includegraphics[width=3.2 in]{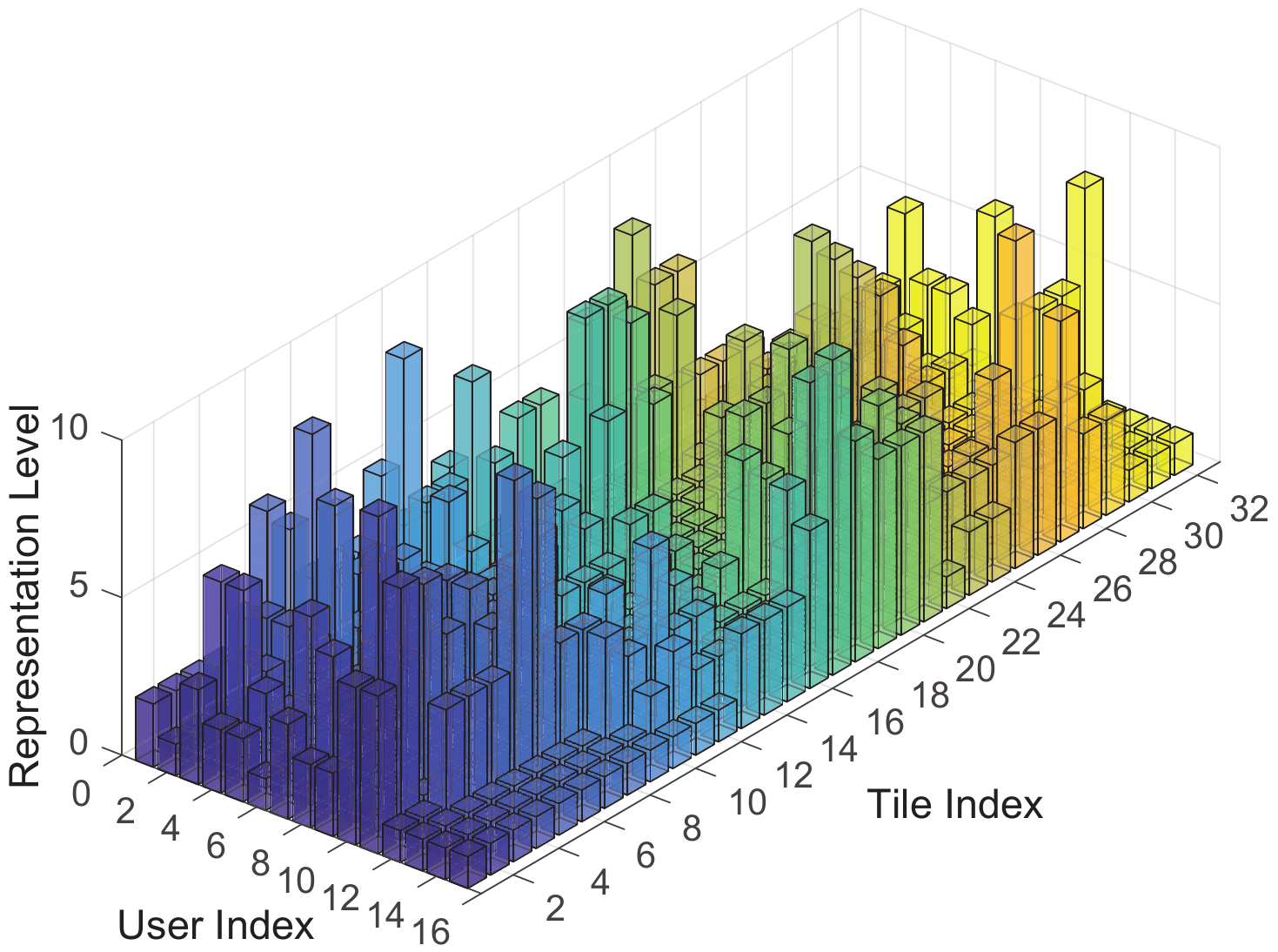}
\caption{Rate allocation of each tile on each user.}\label{HW_figure_Rate allocation}
\end{figure}

\textbf{Impact on Different Users:} Detailed rate allocation on each user is depicted in Figure 7. Although all the saliency is normalized for each 360-degree video, the user can still  look at distinct kinds of views and their behaviors are different. The achievable rates of users 14 and 15 are very close and both are looking at the same video. However, they are looking at different FoVs, therefore user 14 gets higher rate due to the influence of the saliency and FoV as shown in Figure 7.

\textbf{Buffer Strategy:} Figure 8(a) shows the channel fluctuation in the network. In such scenario, we assume that we only have the current channel states information of the network. However, we do not have an estimation of future states. The server adopts buffer strategy always according to the current states. Figure 8(b) demonstrates the utility of one user along the network fluctuation with two benchmark buffer strategies.  We can find that our hierarchical buffer updating strategy performs much better than others. Short-buffer strategy can achieve high quality when the buffer is not empty. However, it is easy to pause and re-buffering when the performance of network becomes bad. Besides, when the network recovers, it takes longer time than our hierarchical buffer updating strategy to get a high utility since there is no updating.

By contrast, long-buffer strategy can avoid the re-buffering events effectively. The current buffer size is always larger than $B_{1}$. Nonetheless, the cost is that it always keeps a low quality (even worse than short-buffer strategy sometimes) since the poor prediction in the buffer's area after $B_{1}$.

Our hierarchical updating strategy can achieve a good tradeoff between the prediction accuracy and the buffer size, as a result, it achieves a high average utility.

\begin{figure}
\centering
\subfigure[]{
\includegraphics[width=3.2 in]{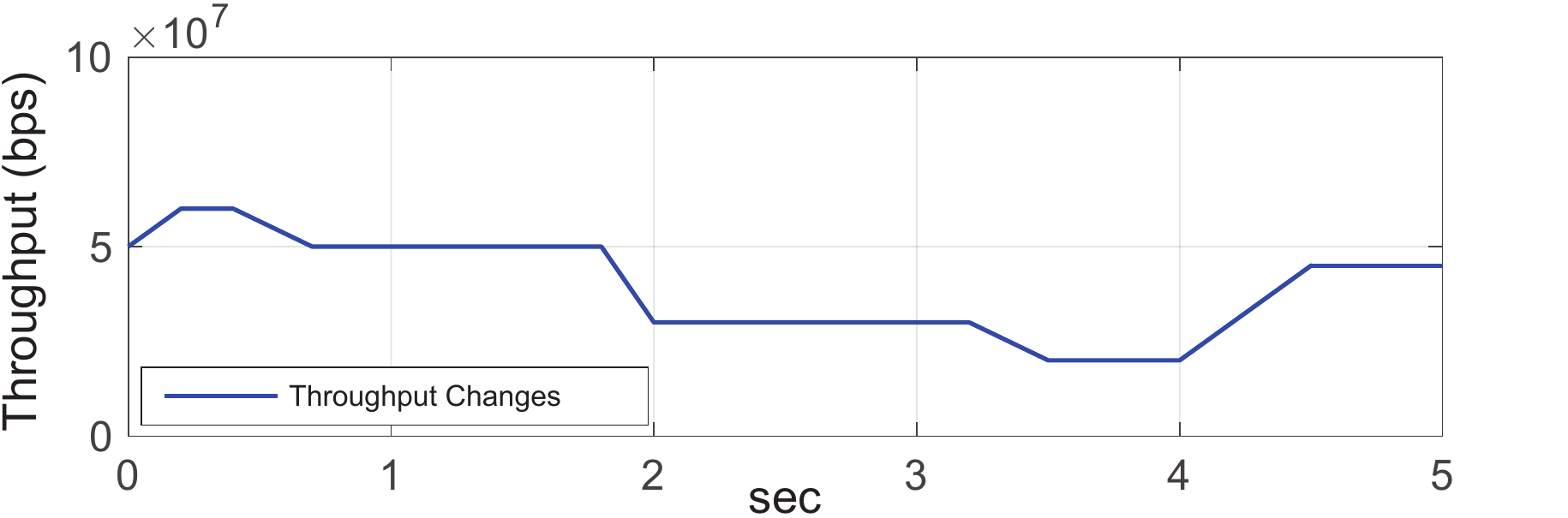}
}
\subfigure[]{
\includegraphics[width=3.2 in]{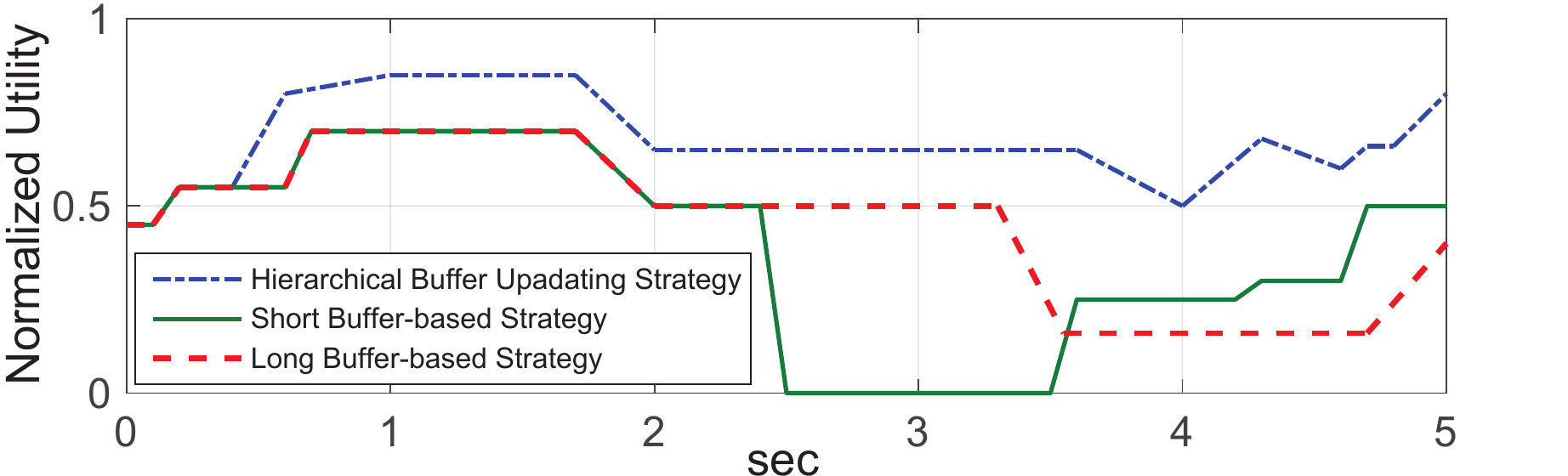}
}
\caption{(a) The time-varying network (b) Utility of different buffer strategies in the time-varying network. }\label{7ab}
\end{figure}

\begin{figure}
\centering
\includegraphics[width=3.2 in]{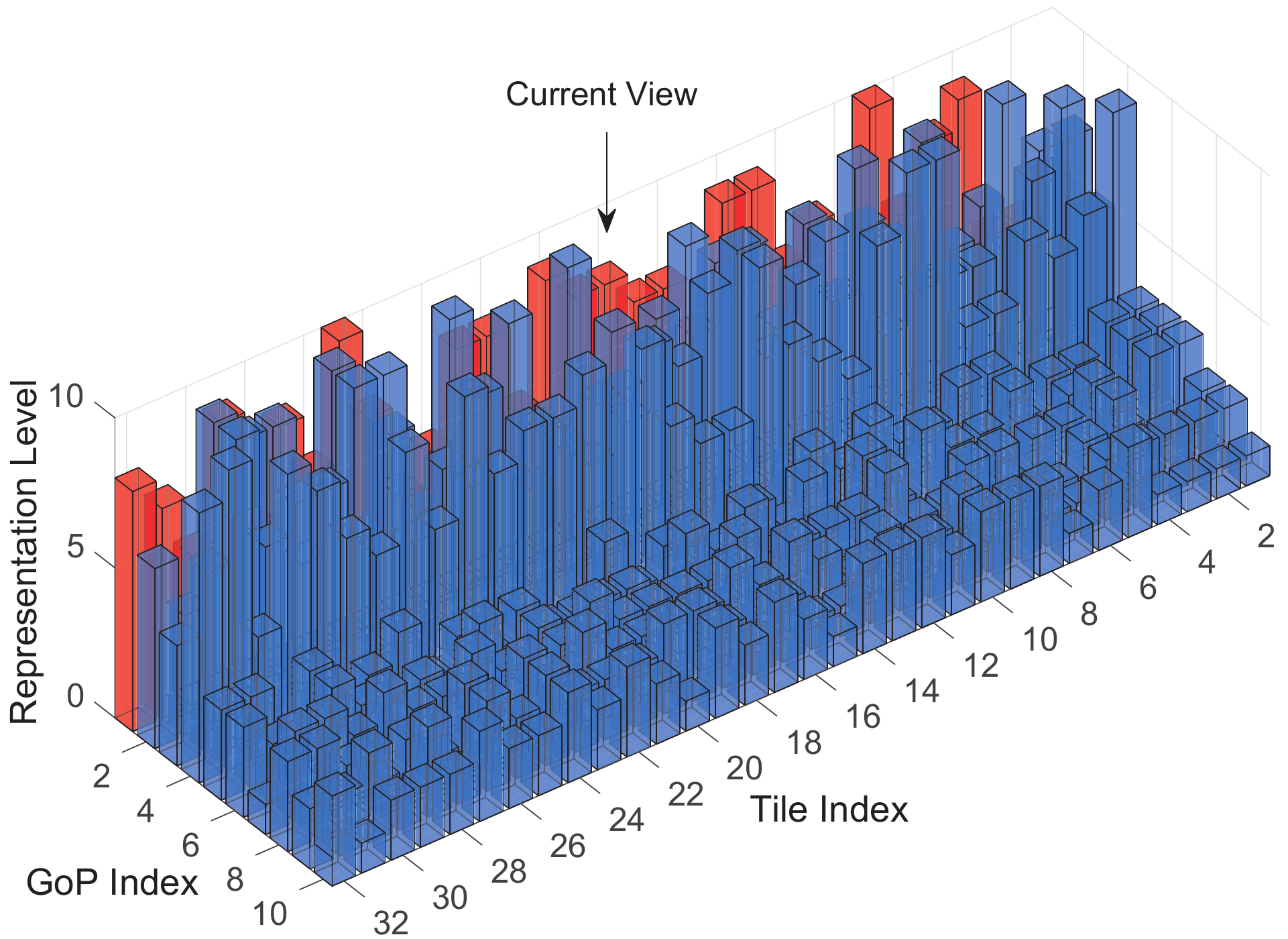}
\caption{Buffer details with hierarchical buffer updating strategy.}\label{9}
\end{figure}

The detailed information of buffering with our hierarchical buffer updating strategy is demonstrated in Figure 9. Here we just show one of the frame in each GoP due to the limit of space. We can find that the levels of representations of some tiles are high in current view and upcoming frames, while it is relatively low after $B_1$. Those high level representation tiles are likely to be FoV according to the high accuracy prediction. Thus, our strategy can make users' FOV a high quality when the bandwidth is enough as well as avoid the buffer being empty when the performance of network turns bad.

\section{Conclusion}

In this paper, we propose a tile-based 360-degree VR video transmission scheme and a corresponding buffer strategy on heterogeneous networks with multi-user access. To better improve the experience of users, we jointly consider saliency in videos, filed of view and the channel quality states of users. The proposed scheme adaptively chooses the most appropriate Wi-Fi AP connection and allocates heterogeneous LTE/WLAN resources at the same time for each tile of each user.  Besides, we proposed a highly effective heuristic search algorithm to solve an NP-hard mixed-integer problem with low complexity. Moreover, a novel buffer updating strategy is proposed to tackle the buffering problem of FoV-driven 360-degree videos. The simulation results show that our proposed scheme and algorithms outperform other methods.

\section*{Acknowledgment}

This work was supported in part by the National Natural Science Foundation of China under Grant 61420106008, Grant 61671301, Grant 61521062, Grant 61650101, in part by the National High Technology Research and Development Program (863 Program) under Grant 2015AA015802.

% use section* for acknowledgment

% Can use something like this to put references on a page
% by themselves when using endfloat and the captionsoff option.
\ifCLASSOPTIONcaptionsoff
  \newpage
\fi

\bibliographystyle{ieeetr}
%\bibliography{Ch_ref}

\begin{thebibliography}{19}

\bibitem{Cisco1}
\newblock ``Cisco visual networking index: Forecast and methodology, 2016$-$2021'', 2016.

\bibitem{Bell}
P. Rondao Alface, M. Jean-Francois, and V. Nico
\newblock ``Interactive omnidirectional video delivery: a bandwidt effective approach,''
\newblock {\em Bell Labs Technical Journal},vol.16, no.4, pp.135-147, 2012


\bibitem{prediction}
Y. Bao, H. Wu, T. Zhang, A.A. Ramli,and X. Liu,
\newblock ``Shooting a moving target: Motion-prediction-based transmission for 360-degree videos,''
\newblock {\em IEEE International Conference on Big Data (Big Data)}, pp.1161-1170, 2016.

\bibitem{arxiv}
A. Ghosh, V. Aggarwal,and F. Qian,
\newblock ``A rate adaptation algorithm for tile-based 360-degree video streaming,''
\newblock {arXiv preprint arXiv:1704.08215.}, 2017.

\bibitem{QEC}
X. Corbillon, G. Simon, A. Devlic, and J. Chakareski,
\newblock ``Viewport-adaptive navigable 360-degree video delivery,''
\newblock {\em IEEE International Conference on Communications (ICC)}, pp. 1-7, 2017.

\bibitem{VR1}
 M. Hosseini and V. Swaminathan,
\newblock ``Adaptive 360 VR video streaming: Divide and conquer,''
\newblock {\em IEEE International Symposium on Multimedia (ISM)}, pp. 107-110, 2016.

\bibitem{FoV}
S. Kim, H. Lee, D. Jeon, and S. Lee,
\newblock ``Reduction in encoding redundancy for overlapped FOVs over wireless visual sensor networks,''
\newblock {\em Digital Signal Processing}, 50, pp. 135-149, 2016.


\bibitem{VR2}
A. Zare, A. Aminlou, M.M. Hannuksela, and M. Gabbouj,
\newblock ``HEVC-compliant tile-based streaming of panoramic video for virtual reality applications,''
\newblock {\em ACM on Multimedia Conference}, pp. 601-605, 2016.



\bibitem{saliency1}
V. Sitzmann, A. Serrano, A. Pavel, M. Agrawala, D. Gutierrez, and G. Wetzstein,
\newblock ``Saliency in VR: How do people explore virtual environments?,''
\newblock {arXiv preprint arXiv:1612.04335}, 2016.

\bibitem{saliency2}
R. Monroy, S. Lutz, T. Chalasani and A. Smolic,
\newblock ``Saliency maps for omni-directional images with CNN,''
\newblock {arXiv preprint arXiv:1709.06505}, 2017.

\bibitem{saliency3}
C. Zhu, K. Huang, and G. Li,
\newblock ``Automatic salient object detection for panoramic images using region growing and fixation prediction model,''
\newblock {arXiv preprint arXiv:1710.04071}, 2017.


\bibitem{motion}
S. Afzal, J. Chen, and K.K. Ramakrishnan,
\newblock ``Characterization of 360-degree Videos,''
\newblock {\em ACM Workshop on Virtual Reality and Augmented Reality Network}, pp.1-6, 2017.


\bibitem{LWA1}
D. Krishnaswamy, D. Zhang, S.Soliman, and B. Mohanty,
\newblock `` Concurrent bandwidth aggregation over wireless networks,''
\newblock {\em IEEE Computing, Networking and Communications (ICNC)}, 2012.


\bibitem{LWA2}
C. Cano, and D.J. Leith,
\newblock ``Coexistence of WiFi and LTE in unlicensed bands: A proportional fair allocation scheme,''
\newblock {\em IEEE Communication Workshop (ICCW)}, 2015.


\bibitem{LWA3}
R. Zhang, M. Wang, L.X Cai,and Z.Zheng,
\newblock ``LTE-unlicensed: the future of spectrum aggregation for cellular networks,''
\newblock {\em IEEE Wireless Communications}, vol.22, no.3, pp.150-159, 2015.


\bibitem{SDN1}
V.G. Nguyen, T.X. Do, and YH. Kim,
\newblock ``SDN and virtualization-based LTE mobile network architectures: A comprehensive survey,''
\newblock {\em Communications Surveys \& Tutorials}, vol.86, no.3, pp.1401-1438, 2016.

\bibitem{SDN2}
S. Kang, W. Yoon,
\newblock ``SDN-based resource allocation for heterogeneous LTE and WLAN multi-radio networks,''
\newblock {\em The Journal of Supercomputing}, vol. 72, no. 4, pp. 1342- 1362, Feb. 2016.

\bibitem{SDNVR1}
S. Mangiante, G. Klas, A. Navon, Z. GuanHua, J. Ran, and M.D. Silva,
\newblock ``VR is on the Edge: How to deliver 360-dgeree videos in mobile networks,''
\newblock {\em ACM Workshop on Virtual Reality and Augmented Reality Network}, pp.30-35, 2016.

\bibitem{SDNVR2}
C. Westphal,
\newblock ``Challenges in networking to support augmented reality and virtual reality,''
\newblock {\em ICNC}, 2017.


\bibitem{SDNVR3}
S. Zhao, and D. Medhi,
\newblock ``SDN-assisted adaptive streaming framework for tile-based immersive content using MPEG-DASH,''
\newblock {researchgate.net}, 2017.


\bibitem{SDNVR4}
E. Bastug, M. Bennis, M. M¨¦dard, and M. Debbah,
\newblock ``Toward interconnected virtual reality: Opportunities, challenges, and enablers,''
\newblock {\em IEEE Communications Magazine}, vol.55, no.6, pp.110-117, 2017.


\bibitem{buffer1}
J. Jiang, V. Sekar, and H. Zhang,
\newblock ``Improving fairness, efficiency, and stability in http-based adaptive video streaming with festive,''
\newblock {\em ACM on Emerging networking experiments and technologies}, pp.97-108, 2012.


\bibitem{buffer2}
L. D. Cicco, V. Caldaralo, V. Palmisano, and S. Mascolo,,
\newblock ``Elastic: A client-side controller for dynamic adaptive streaming over http (DASH),''
\newblock {\em IEEE Packet Video Workshop (PV)}, pp.1-8, 2013.


\bibitem{buffer3}
Z. Li, A.C. Begen, J. Gahm, Y. Shan, B. Osler, and D. Oran,
\newblock ``Streaming video over HTTP with consistent quality,''
\newblock {\em ACM Multimedia Systems Conference}, pp.248-258, 2014.

\bibitem{re-buffering}
T. D. Pessemier, K. D. Moor, W. Joseph, L. D. Marez and L. Martens
\newblock ``Quantifying the influence of rebuffering interruptions on the user's quality of experience during mobile video watching,''
\newblock {\em IEEE Transactions on Broadcasting}, vol.59, no.1, pp.47-61, 2013.


\bibitem{S1}
A. Papushoy, and A. G. Bors,
\newblock ``Image retrieval based on query by saliency content,''
\newblock {\em  Digital Signal Processing}, 36, pp.156-173. 2015.


\bibitem{S2}
M. Du, X. Wu, W. Chen, and Z. Li,
\newblock ``Supervised training and contextually guided salient object detection,''
\newblock {\em  Digital Signal Processing}, 63, pp.44-55. 2017.



\bibitem{saliency4}
U. Engelke, M. Barkowsky, P. Le Callet, and H. J. Zepernick,
\newblock ``Modelling saliency awareness for objective video quality assessment,''
\newblock {IEEE In Quality of Multimedia Experience (QoMEX)}, pp. 212-217, 2010.


\bibitem{QoE4}
A. Aldahdooh, E. Masala, G. Van Wallendael, and M. Barkowsky,
\newblock ``Framework for reproducible objective video quality research with case study on PSNR implementations,''
\newblock {\em Digital Signal Processing}, 2017.



\bibitem{QoE0}
Y. Liu, J. Yang, Q. Meng, Z. Lv, Z. Song, and Z. Gao,
\newblock ``Stereoscopic image quality assessment method based on binocular combination saliency model,''
\newblock {\em Signal Processing}, 125, pp.237-248, 2016.



\bibitem{QoE}
F. Shao, G. Y. Jiang, M. Yu, F. Li, Z. Peng, and R. Fu,
\newblock ``Binocular energy response based quality assessment of stereoscopic images,''
\newblock {\em Digital Signal Processing}, 29, pp.45-53, 2014.



\bibitem{QoE2}
P. Reichl, S. Egger, R. Schatz, and A. D'Alconzo,
\newblock ``The logarithmic nature of QoE and the role of the Weber-Fechner law in QoE assessment,''
\newblock {\em IEEE International Conference on Communications (ICC)},2010.


\bibitem{QoE3}
L. Qian, Z. Cheng, Z. Fang, L. Ding, F. Yang and W. Huang
\newblock ``A QoE-driven encoder adaptation scheme for multi-user video streaming in wireless networks,''
\newblock {\em IEEE Transactions on Broadcasting}, vol. 63, no. 1, pp.20-31, 2017.



\bibitem{QoE1}
W. Zhang, Y. Wen, Z. Chen, and A. Khisti,
\newblock ``QoE-Driven cache management for HTTP adaptive bit rate streaming over wireless networks,''
\newblock {\em IEEE Transactions on Multimedia }, vol. 15, no. 6, pp.1431-1445, 2013.

\bibitem{MPD}
T. Stockhammer,
\newblock ``Dynamic adaptive streaming over HTTP,''
\newblock {\em ISO/IEC, MPEG Draft International Standard}, 2011.


\bibitem{representation}
L. Yu, T. Tillo, and J. Xiao,
\newblock ``QoE-driven dynamic adaptive video streaming strategy with future information,''
\newblock {\em IEEE Transactions on Broadcasting}, vol. 61, no. 4, pp.651-665, 2015.


\bibitem{feedback1}
L. Song, Z. Han, Z. Zhang, and B. Jiao,
\newblock ``Non-cooperative feedback-rate control game for channel state information in wireless networks,''
\newblock {\em IEEE Journal on Selected Areas in Communications}, vol. 30, no. 1, pp.188-197, 2012.


\bibitem{feedback2}
Q. Xu, S. Mehrotra, Z. Mao, and J. Li,
\newblock ``PROTEUS: network performance forecast for real-time, interactive mobile applications,''
\newblock {\em In Proceeding of the 11th annual international conference on Mobile systems}, ACM, pp. 347-360, 2013.


\bibitem{prediction1}
F. Qian, L. Ji, B. Han, and V. Gopalakrishnan,
\newblock ``Optimizing 360 video delivery over cellular networks,''
\newblock {\em In Proceedings of the 5th Workshop on All Things Cellular: Operations, Applications and Challenges}, ACM, pp. 1-6, 2016.


\bibitem{prediction2}
C.L. Fan, J. Lee, W.C. Lo, C.Y. Huang, K.T. Chen, and C.H. Hsu,
\newblock ``Fixation prediction for 360 video streaming in head-mounted virtual reality,''
\newblock {\em In Proceedings of the 27th Workshop on Network and Operating Systems Support for Digital Audio and Video}, ACM, pp. 67-72, 2017.

\bibitem{projection}
\newblock ``Information technology - Coded representation of immersive media (MPEG-I) - Part 2: Omnidirectional media format''
\newblock {\em ISO/IEC JTC1/SC29/WG11  MPEG Draft International Standard}, 2017.


\bibitem{TSP}
K. L. Hoffman, M. Padberg, and G. Rinaldi,
\newblock ``Traveling salesman problem,''
\newblock {\em In Encyclopedia of operations research and management science,} Springer US, pp. 1573-1578, 2013.



\bibitem{convex1}
S. Boyd and L. Vanderberghe,
\newblock `` Convex Optimization,''
\newblock {\em Cambridge university press}, 2004.


\bibitem{convex2}
M. Grant, S. Boyd and Y. Ye
\newblock ``Convex Optimization,''
\newblock {CVX: MATLAB software for disciplined convex programming}, 2008

\bibitem{buff4}
T.Y. Huang, R. Johari, N. McKeown, M. Trunnell and M. Watson,
\newblock ``A buffer-based approach to rate adaptation: Evidence from a large video streaming service,''
\newblock {\em  ACM SIGCOMM Computer Communication Review}, vol.44, no.4, pp.187-198, 2015.


\bibitem{buff5}
K. Spiteri, R. Urgaonkar, and R.K. Sitaraman,
\newblock ``BOLA: near-optimal bitrate adaptation for online videos,''
\newblock {\em IEEE INFOCOM 2016-The 35th Annual}, pp.1-9, 2016.

\bibitem{SRD}
O.A. Niamut, E. Thomas, L. D'Acunto, C. Concolato, F. Denoual and S.Y. Lim,
\newblock ``MPEG DASH srd: Spatial relationship description,''
\newblock {\em In Proceedings of the 7th International Conference on Multimedia Systems}, ACM, pp.5, 2016.


\end{thebibliography}

\end{document}